\documentclass[prb,twocolumn,showpacs,amsmath,amssymb,superscriptaddress]{revtex4}
\pdfoutput=1

\usepackage{epsfig}
\usepackage{ graphicx } % Include figure files
\usepackage{ bm } % bold math
\usepackage{ color }

\newcommand{\ua}{\uparrow}
\newcommand{\da}{\downarrow}

\newcommand{\be}{\begin{equation}}
\newcommand{\ee}{\end{equation}}
\newcommand{\bea}{\begin{eqnarray}}
\newcommand{\eea}{\end{eqnarray}}

\newcommand{\veps}{\varepsilon}
\newcommand{\vk}{{\boldsymbol k}}

\begin{document}
\title{Fate of Majorana fermions and Chern numbers after a quantum quench}
\author{ P.D. Sacramento }
%\email{ pdss@cfif.ist.utl.pt }
\affiliation{ \textit Centro de F\'isica das Intera\c c\~oes Fundamentais, 
Instituto Superior T\'ecnico, Universidade de Lisboa, Av. Rovisco Pais, 1049-001 Lisboa, Portugal }
\date{ \today }

%%%%%%%%%%%%%%%%%%%%       ABSTRACT      %%%%%%%%%%%%%%%%%%%%

\begin{abstract}
The stability of Majorana fermions at the edges of a two-dimensional
topological supercondutor is studied, after quenches to either non-topological phases or
other topological phases. Both instantaneous and slow quenches are considered. In general, 
the Majorana modes decay and, in the case of instantaneous quenches, their revival times
scale to infinity as the system size grows. Considering fast quantum quenches within the
same topological phase, leads, in some cases, to robust edge modes. Quenches to a
topological $Z_2$ phase reveal some robustness of the Majorana fermions. Comparing strong
spin-orbit coupling with weak spin-orbit coupling, it is found that the Majorana fermions
are fairly robust, if the pairing is not aligned with the spin-orbit Rashba coupling.
It is also shown that the Chern number remains invariant after the quench, until the
propagation of the mode along the transverse direction reaches the middle
point, beyond which the Chern number oscillates between increasing values. In some cases,
the time average Chern number seems to converge to the appropriate value, but often the
decay is very slow. The effect of varying the rate of change in slow quenches is also
analysed. It is found that the defect production is non-universal and does not follow the Kibble-Zurek scaling
with the quench rate, as obtained before for other systems with topological edge states. 
\end{abstract}

\pacs{05.30.Rt, 05.70.Ln, 03.65.Vf}

\maketitle

\section{Introduction}

The time evolution of a quantum system coupled to a dissipative environment
has attracted interest for a long time \cite{leggett} and, in particular, the
problem of thermalization associated with the coupling to a heat bath.
Sudden quenches, associated with abrupt changes of some external parameters,
such as magnetic fields or temperature,
and in general discontinuous phase transitions, have been studied in
various contexts, both in classical \cite{gunton} and quantum systems, 
involving, in general, the formation and growth of a seed
of a stable phase, inside a metastable or unstable phase, such as
in a spinodal decomposition \cite{spinodal}. An example of theoretical and
practical interest is the growth of magnetic bubbles in magnetic systems
away from equilibrium \cite{bubbles}.

An abrupt change of the state of an isolated quantum system leads to a unitary
time evolution and, therefore, the issue of thermalization raises interesting
questions \cite{polkovnikov1}. The end state of this evolution has attracted interest
due to the prediction of different outcomes depending on the type of
system and have been confirmed by recent experiments \cite{exp}. 
In general, it is expected some sort of thermalization in the sense that
correlation functions stabilize, in a way similar to a statistical
description, at some effective temperature \cite{thermal}. 
This convergence is explained in terms of the hypothesis of
eigenstate thermalization that occurs at the level of each eigenstate \cite{deutsch,rigol1}.
Interesting
exceptions are soluble and integrable systems where thermalization breaks down as one
approaches an integrable point. However, some sort of 
thermalization is predicted, such as the one observed in integrable
systems, for which an equilibrium like distribution is expected in terms of
a generalized Gibbs ensemble, of the (infinitely) many conserved
quantities \cite{rigol2,cazalilla,rigol3,luttinger}.
A possible way to produce such a change is performing a quantum
quench obtained changing abruptly the Hamiltonian parameters.

Slow transitions are qualitatively different.
In the field of thermal transitions crossing a critical point involves a change between states with
different symmetries. Close to the critical point fluctuations are able to
sample domains of the most stable phase and lead to some dynamical scaling, in terms of the rate
of change of a driving parameter across the transition,
as proposed by Kibble and Zurek \cite{kibble,zurek1}.  
A similar behavior is expected around a quantum critical point \cite{zurek2,polkovnikov2}.
In both cases the transition induces a density of excitations that scales with the transition
rate with some critical exponent.

On the other hand, topological systems have attracted interest \cite{kane,zhang} and, specifically,
topological superconductors \cite{alicea} due to the prediction of Majorana fermions.
Their interest is twofold: first as a physical realization of long sought-after
Majorana fermions, and second as possible elements in quantum computation, due to their
non-abelian statistics when combined with vortices or other local entities.
Topological systems are intrinsically interesting due to their robust properties,
and efforts towards the understanding of their properties have attracted
interest, and in particular, a great effort has been put towards prediction and
detection of Majorana fermions \cite{mourik}.

Their robustness is a key property. It is therefore interesting to study their
robustness to various perturbations. In particular, it is interesting to study
their response to time dependent perturbations and, in particular, a quantum quench.
The presence of a nontrivial topological phase is frequently associated to other
phases, as some parameter or parameters in the Hamiltonian change.
These changes may lead to closing of energy gaps in the spectrum that may originate
a transition to a trivial phase, or to some other phase characterized by a different
topology. It has been shown before that topological systems are quite robust to a 
quantum quench, as exemplified by the toric code model \cite{tsomokos,hamma}. 
With appropriate boundary conditions topological systems show gapless edge states, such as
the Majorana fermions in topological superconductors.
It is therefore interesting to determine their stability to a quantum quench. 

Examples that host Majorana fermions as edge states are provided by several
superconducting systems, such as a one-dimensional fully polarized p-wave superconductor
($1d$ Kitaev model \cite{kitaev}), two-dimensional triplet $p+ip$ superconductor \cite{pip} and various
other systems that mix superconducting order (eventually by proximity effects)
with Zeeman fields and/or spin-orbit coupling \cite{sato}. Here we will focus attention
on a two-dimensional triplet superconductor with Zeeman field, $M_z$, and Rashba spin orbit
coupling, that has been shown to have a rich phase diagram with various topological
and trivial phases \cite{sato,epl}.

A quantum quench between different points in the phase diagram leads to a time evolution
between a system characterized by some sort of topological order to another phase that may
be trivial or non-trivial.
The effect of quantum quenches on nontrivial edge states was carried out before
considering both slow rates and fast quenches. In the context 
of the Creutz ladder, it was shown that the presence of edge states modifies the
process of defect production expected from the Kibble-Zurek mechanism, leading
in this problem to a scaling with the change rate with a non-universal critical
exponent \cite{bermudez1}. A similar result was obtained for the one-dimensional
superconducting Kitaev model, where it was shown that, although bulk states follow
the Kibble-Zurek scaling, the produced defects for an edge state quench are quite
anomalous and independent of the quench rate \cite{bermudez2}. 

The behavior of edge states under an abrupt quantum quench has also been considered
very recently in the context of a two-dimensional topological insulator \cite{bhz},
where it was found that, in the sudden transition from the topological insulator to
the trivial insulator phase, there is a collapse and revival of the edge states \cite{patel}.
Similar results were obtained for the one-dimensional Kitaev model \cite{rajak}, also
studying the signature of the Majoranas in the entanglement spectrum \cite{chung}.

In this work we will focus
attention on the time evolution of a Majorana fermion, characteristic of
a nontrivial phase on a two-dimensional triplet superconductor with Zeeman field, $M_z$, and Rashba spin orbit
coupling, as a quench is performed. In particular, in the case of a fast quench the survival probability
of such a state is studied, and it is found that its robustness is in general lost,
except for some particular cases. Also, the time evolution of the Chern number is studied
across the transition. Slow quenches are also considered and a non-universal behavior is
found in agreement with other topological edge states. 
In sections II and III a brief review of a quantum quench and the triplet superconductor is
presented. In section IV we present results for a finite system using a real space description,
and compare results for the one-dimensional Kitaev model and the two-dimensional superconductor, and stress
the influence of the spin-orbit coupling on the robustness of the edge states, after an
abrupt quench. 
In section V  we study the stability of the edge states in momentum space, and in section VI
the evolution of the Chern numbers. In section VII slow quenches are considered from the
regime of quasi-adiabatic transitions to fast quenches. We conclude with section VIII.

\section{Quantum quenches}

Let us label a quantum state of the system at some initial state as
$|\psi_m(\xi)\rangle$, where $\xi$ represents a set of parameters upon
which the Hamiltonian depends. 
Here it may be the set $\xi=(M_z,\epsilon_F)$, where $\epsilon_F$ is the
chemical potential.
As the parameters change, the system goes through different topological phases.
We will consider that at the initial time, $t=0$, an abrupt change of the parameters is performed
to some set $\xi^{\prime}$. After this sudden quench
the system will evolve in time under the influence of a different Hamiltonian.
Since the Hamiltonian is quadratic, its eigenstates are easilly obtained solving
for the single particle modes, given by the solution of the BdG equations.
Since the Hamiltonian changes, the initial states, (calculated at $t=0^-$), are
no longer eigenstates, will mix and evolve in time as
\be
|\psi_m(\xi,t)\rangle = e^{-iH(\xi^{\prime})t} |\psi_m(\xi)\rangle
\label{evolve}
\ee
Denoting the eigenstates of the Hamiltonian, $H(\xi^{\prime})$, as $|\psi_n(\xi^{\prime})\rangle$ and the
eigenvalues as $E_n(\xi^{\prime})$, we may write that
the time evolved state is given by
\be
|\psi_m(\xi,t)\rangle = \sum_n e^{-i E_n(\xi^{\prime})t} |\psi_n(\xi^{\prime})\rangle
\langle \psi_n(\xi^{\prime})|\psi_m(\xi)\rangle
\ee

Moreover, we may calculate the overlap of this time evolved state with the initial state
leading to
\be
A_m(t)=\langle \psi_m(\xi)|\psi_m(\xi,t)\rangle
\ee
This can be expressed as
\be
A_m(t)=\sum_n | \langle \psi_m(\xi)|\psi_n(\xi^{\prime})\rangle|^2 e^{-iE_n(\xi^{\prime})t}
\ee
and involves the overlap between the initial state and all the eigenstates of the
final Hamiltonian. Following \cite{rajak} we may define a survival probability for the
initial state as
\be
P_m(t)=|A_m(t)|^2
\ee

In this work we will be interested in the fate of single particle states after a quantum
quench across the phase diagram. We consider a subspace of one excitation such that
the total Hamiltonian is given by the ground state energy plus one excited state.
We will consider lowest energy excitations which, in the nontrivial topological
phases, are Majorana fermions. We will take the initial state $|\psi_m(\xi)\rangle$ as one
of these lowest energy states (eigenstate of the single particle Hamiltonian for the set
of parameters $\xi$). Since we remain in the one excitation subspace after the quench, the
Hamiltonian that gives the unitary time evolution is the single particle Hamiltonian for
the set of parameters $\xi^{\prime}$ and the eigenstates, $|\psi_n(\xi^{\prime})\rangle$, are
the single particle states of the new Hamiltonian. We will consider later two descriptions
of the superconductor. First we will consider a finite system, with periodic boundary conditions
along one space direction, say $x$, and open boundary conditions along the other space direction, say $y$.
This leads to the appearance of edge states along the $x$-direction edges. Next, we will use a
momentum space representation of the states along $x$ labelling the states by the coordinate $y$ and
the momentum $k_x$. Since the Hamiltonian is diagonal in momentum space, it is enough to consider
the overlap between the various eigenstates within each momentum value, $k_x$. 
In the real space description the single particle states involved are defined over the entire
system.

\begin{figure}[t]
\includegraphics[width=0.95\columnwidth]{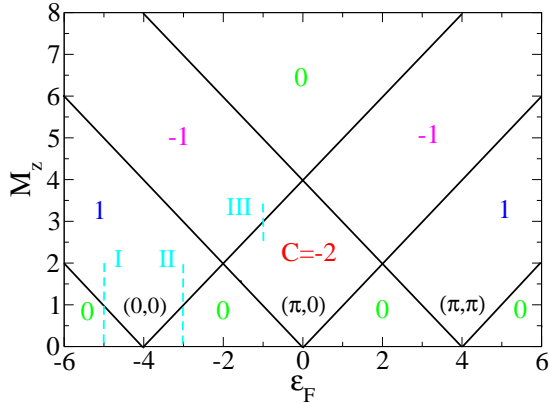}
\caption{\label{fig1}
(Color online) Topological phases and their Chern (C) numbers as a function of chemical potential and magnetization.
The transitions occur at three sets of momenta $\boldsymbol{k}=(0,0),\boldsymbol{k}=(\pi,0),\boldsymbol{k}=(\pi,\pi)$.
The three dashed lines $I,II,III$ correspond to the slow quenches considered in section VII.}
\end{figure}

\section{Two-dimensional triplet superconductor}

We consider a two-dimensional triplet superconductor
with $p$-wave symmetry.
This model was studied in Refs. \onlinecite{sato,epl}.
We write the Hamiltonian for the bulk system as
\begin{eqnarray}
\hat H = \frac 1 2\sum_\vk  \left( {\boldsymbol \psi}_{\vk}^\dagger ,{\boldsymbol \psi}_{-\vk}   \right)
\left(\begin{array}{cc}
\hat H_0(\vk) & \hat \Delta(\vk) \\
\hat \Delta^{\dagger}(\vk) & -\hat H_0^T(-\vk) \end{array}\right)
\left( \begin{array}{c}
 {\boldsymbol \psi}_{\vk} \\  {\boldsymbol \psi}_{-\vk}^\dagger  \end{array}
\right)
\label{bdg1}
\end{eqnarray}
where $\left( {\boldsymbol \psi}_{\vk}^{\dagger}, {\boldsymbol \psi}_{-\vk} \right) =
\left( \psi_{\vk\ua}^{\dagger}, \psi_{\vk\da}^\dagger ,\psi_{-\vk\ua}, \psi_{-\vk\da}   \right)$
and
\begin{equation}
\hat H_0=\epsilon_\vk\sigma_0 -M_z\sigma_z + \hat H_R\,.
\end{equation}
Here, $\epsilon_{\boldsymbol{k}}=-2 \tilde{t} (\cos k_x + \cos k_y )-\veps_F$
is the kinetic part, $\tilde{t}$ denotes the hopping parameter set in
the following as the energy scale, $\veps_F$ is the
chemical potential,
$\boldsymbol{k}$ is a wave vector in the $xy$ plane, and we have taken
the lattice constant to be unity. Furthermore, $M_z$
is the Zeeman splitting term responsible for the magnetization,
in energy units.
The Rashba spin-orbit term is written as
\begin{equation}
\hat H_R = \boldsymbol{s} \cdot \boldsymbol{\sigma} = \alpha
\left( \sin k_y \sigma_x - \sin k_x \sigma_y \right)\,,
\end{equation}
 where
$\alpha$ is measured in the energy units
 and $\boldsymbol{s} =\alpha(\sin k_y,-\sin k_x, 0)$.
The matrices $\sigma_x,\sigma_y,\sigma_z$ are
the Pauli matrices acting on the spin sector, and $\sigma_0$ is the
$2\times 2$ identity.
The pairing matrix reads
\begin{equation}
\hat \Delta = i\left( {\boldsymbol d}\cdot {\boldsymbol\sigma} \right) \sigma_y =
 \left(\begin{array}{cc}
-d_x+i d_y & d_z \\
d_z & d_x +i d_y
\end{array}\right)\,.
\end{equation}
The pairing matrix for a p-wave superconductor generally satisfies
$\hat \Delta \hat \Delta^{\dagger} = |\boldsymbol d|^2 \sigma_0 + \boldsymbol q \cdot \boldsymbol{\sigma}\,$,
where $\boldsymbol q=i \boldsymbol d \times \boldsymbol d^*$.
If the vector $\boldsymbol q$ vanishes the pairing is called unitary. 
Otherwise it is called non-unitary \cite{Sigrist} and breaks time-reversal symmetry (TRS),
originating a spontaneous magnetization in the system due to the symmetry of the pairing, as in $^3He$.
We will consider unitary pairing. 
If the spin-orbit is strong the pairing is aligned \cite{Sigrist2} along the
spin-orbit vector $\boldsymbol{s}$. 
This case is denoted by strong coupling case. 
Relaxing this restriction allows that the two vectors are not aligned. This case
is denoted by weak spin-orbit coupling.
In the strong-coupling case $\boldsymbol{d}=(d_x,d_y,d_z) = ( d / \alpha ) \boldsymbol{s} $
and $d$ is a scale parameter.
As an example of the weak coupling pairing we will take 
$d_x=d \sin k_y, d_y=d \sin k_x, d_z=0$
(other cases were considered in \cite{ahe,epl}).

The energy eigenvalues and eigenfunction may be obtained solving the Bogoliubov-de Gennes equations
\be
\label{bdg2}
\left(\begin{array}{cc}
\hat H_0(\vk) & \hat \Delta(\vk) \\
\hat \Delta^{\dagger}(\vk) & -\hat H_0^T(-\vk) \end{array}\right)
\left(\begin{array}{c}
u_n\\
v_n
\end{array}\right)
= \epsilon_{\boldsymbol{k},n}
\left(\begin{array}{c}
u_n\\
v_n
\end{array}\right).
\ee
The 4-component spinor can be written as
\be
\left(\begin{array}{c}
u_n\\
v_n
\end{array}\right)=
\left(\begin{array}{c}
u_n(\boldsymbol{k},\uparrow) \\
u_n(\boldsymbol{k},\downarrow) \\
v_n(-\boldsymbol{k},\uparrow) \\
v_n(-\boldsymbol{k},\downarrow) \\
\end{array}\right) .\ee

One way to characterize various topological phases is through the Chern number,
obtainable as an integral over the Brillouin zone of the Berry curvature \cite{xiao,Fukui}.
Summing over the occupied bands the Chern number has been calculated \cite{sato,epl}.
The results in the parameter space are shown in Fig. \ref{fig1}
using the typical parameters $\tilde{t}=1$, $\alpha=0.6$, $d=0.6$.

The superconductor we consider here
is time-reversal invariant if the Zeeman term is absent.
The system then belongs to the symmetry class DIII where the topological invariant is
a $\mathbb{Z}_2$ index.
If the Zeeman term is finite, TRS is broken and the system belongs
to the symmetry class D.
The topological invariant that characterizes this phase is the first Chern number $C$,
and the system is said to be a $\mathbb{Z}$~topological superconductor.

\section{Real space description}

We consider first a finite system of dimensions $N_x \times N_y$ along a longitudinal, $x$,
direction and a transversal direction along $y$,
we apply periodic boundary conditions along the $x$ direction and open boundary conditions
along the transverse direction.
We write
\be
\psi_{k_x,k_y,\sigma} = \frac{1}{\sqrt{N_y}} \sum_{j_y} e^{-i k_y j_y} \frac{1}{\sqrt{N_x}} \sum_{j_x} e^{-i k_x j_x}
 \psi_{j_x,j_y,\sigma}\,,
\label{operators2}
\ee
and rewrite the Hamiltonian matrix in terms of
the operators (\ref{operators2})  as
\bea
H = \sum_{j_x} \sum_{j_y}
& & \left(\begin{array}{cccc}
\psi_{j_x,j_y,\uparrow}^{\dagger}  & \psi_{j_x,j_y,\downarrow}^{\dagger} &
\psi_{j_x,j_y,\uparrow}  & \psi_{j_x,j_y,\downarrow}
\end{array}\right) \nonumber \\
& & \hat{H}_{j_x,j_y}
\left(\begin{array}{c}
\psi_{j_x,j_y,\uparrow} \\
\psi_{j_x,j_y,\downarrow} \\
\psi_{j_x,j_y,\uparrow}^{\dagger} \\
\psi_{j_x,j_y,\downarrow}^{\dagger} \\
\end{array}\right)
\eea

The operator $\hat{H}_{j_x,j_y}$ reads
\be
\hat{H}_{j_x,j_y}=
\left(\begin{array}{cc}
A & B \\
C & D \\
\end{array}\right)
\ee
where
\be
A=
\left(\begin{array}{cc}
-M_z-\epsilon_F-\tilde{t} \eta_+^x -\tilde{t} \eta_+^y & \frac{\alpha}{2} \eta_-^x +\frac{\alpha}{2i} \eta_-^y \\
-\frac{\alpha}{2} \eta_-^x +\frac{\alpha}{2i} \eta_-^y & M_z -\epsilon_F -\tilde{t} \eta_+^x -\tilde{t} \eta_+^y \\
\end{array}\right)
\ee
\be
B=
\left(\begin{array}{cc}
-\frac{d}{2} \eta_-^x -\frac{d}{2i} \eta_-^y & 0 \\
0 & -\frac{d}{2}\eta_-^x +\frac{d}{2i} \eta_-^y \\
\end{array}\right)
\ee
\be
C=
\left(\begin{array}{cc}
\frac{d}{2} \eta_-^x -\frac{d}{2i} \eta_-^y & 0 \\
0 & \frac{\alpha}{2} \eta_-^x +\frac{d}{2i} \eta_-^y \\
\end{array}\right)
\ee
\be
D=
\left(\begin{array}{cc}
M_z+\epsilon_F+\tilde{t} \eta_+^x + \tilde{t} \eta_+^y & -\frac{\alpha}{2} \eta_-^x +\frac{\alpha}{2i} \eta_-^y \\
\frac{\alpha}{2} \eta_-^x +\frac{\alpha}{2i} \eta_-^y & -M_z +\epsilon_F +\tilde{t} \eta_+^x +\tilde{t} \eta_+^y \\
\end{array}\right)
\ee
where $\psi_{j_x,j_y}^{\dagger} \eta_{\pm}^x \psi_{j_x,j_y} = 
\psi_{j_x,j_y}^{\dagger} \psi_{j_x+1,j_y} \pm \psi_{j_x+1,j_y}^{\dagger} \psi_{j_x,j_y}$.
and $\psi_{j_x,j_y}^{\dagger} \eta_{\pm}^y \psi_{j_x,j_y} = 
\psi_{j_x,j_y}^{\dagger} \psi_{j_x,j_y+1} \pm \psi_{j_x,j_y+1}^{\dagger} \psi_{j_x,j_y}$.
The diagonalization of this Hamiltonian involves the solution of a $(4 N_x N_y) \times (4 N_x N_y)$ eigenvalue problem.
The energy states include states in the bulk and states along the edges.

\subsection{$1d$ Kitaev model}

\begin{figure}[t]
\includegraphics[width=0.99\columnwidth]{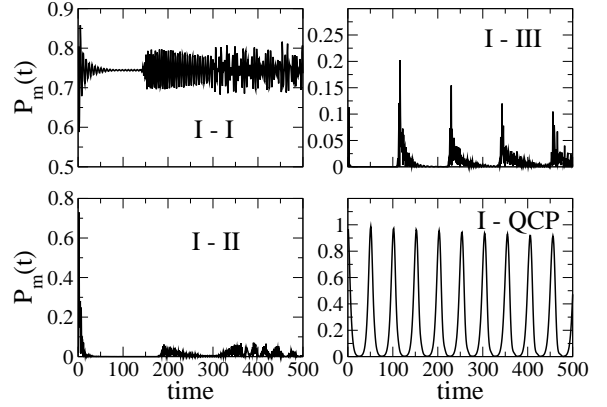}
\caption{\label{fig2}
(Color online) 
Survival probability of the Majorana state of the one-dimensional fully polarized p-wave
Kitaev model for different transitions across the phase diagram: 
i) transition within
the same topological phase, $I$, ($\epsilon_F=0.5,\Delta=0.6) \rightarrow (\epsilon_F=1.0,\Delta=0.6)$,
ii) transition from the topological phase $I$ to the trivial phase $III$
($\epsilon_F=0.5,\Delta=0.6) \rightarrow (\epsilon_F=2.2,\Delta=0.6)$,
iii) transition from the topological phase $I$ with positive $\Delta$ to the topological phase $I$
with negative $\Delta$
($\epsilon_F=0.5,\Delta=0.6) \rightarrow (\epsilon_F=0.5,\Delta=-0.6)$,
iv) transition within
the same topological phase, $I$, to the quantum critical point ($\epsilon_F=0,\Delta=0.1) \rightarrow (\epsilon_F=0,\Delta=0)$.
The system has $100$ sites.
}
\end{figure}

Consider first a one-dimensional spinless p-wave superconductor. Kitaev's model can be written as
\bea
H &=& -\tilde{t} \sum_i \left( c_i^{\dagger}  c_{i+1} + c_{i+1}^{\dagger} c_i \right) -\epsilon_F \sum_i 
\left(c_i^{\dagger} c_i -\frac{1}{2} \right)
\nonumber \\
&+& \Delta \sum_i \left( c_i c_{i+1} + c_{i+1}^{\dagger} c_i^{\dagger} \right)
\eea
The BdG equations for the wave functions may be written as
\be
H\left(\begin{array}{c} u_n(i) \\ v_n(i) \end{array}\right)
= \epsilon_n
\left(\begin{array}{c} u_n(i) \\ v_n(i) \\
\end{array}\right)
\ee
where
\be
H=
\left(\begin{array}{cc}
-\tilde{t}(s_1+s_{-1})-\epsilon_F  & -\Delta (s_i-s_{-1} \\
\Delta (s_1-s_{-1}) & \tilde{t}(s_1+s_{-1} +\epsilon_F \\
\end{array}\right)
\ee
where $s_{\pm 1} f(i)=f(i\pm 1)$ for any function of the lattice point $i$.
The solution of these equations involves the diagonalization of a $2N \times 2N$ matrix,
where $N$ is the number of sites of the superconductor and where open boundary conditions
are used.

The stability of the Majorana fermions in this model has been considered recently
\cite{rajak}. We present here some results as a preview of the results for the two-dimensional
superconductor. In Fig. \ref{fig2} we present results for the survival probability of the
Majorana mode for several quenches. The phase diagram may be found, for instance in the same
reference. There are two topological phases, $I$ and $II$ and two trivial phases denoted $III$.
In the first panel we consider the case of a quench within the same topological phase, in this
instance inside phase $I$. It is clearly shown that the survival probability is finite.
Since the parameters change, the survival probability is not unity, there is a decrease as
a function of time due to the overlap with {\it all} the eigenstates of the chain with the new
set of parameters, but after some oscillations the survival rate stabilizes at some finite
value. As time grows, oscillations appear again centered around some finite value. 
Therefore the Majorana mode is robust to the quench.
In the second panel we consider a quench from the topological phase $I$ to the trivial, non-topological
phase $III$. The behavior is quite different. After the quench the survival probability decays fast to nearly zero.
After some time it increases sharply and repeats the decay and revival process. 
Similar results are found for a quench between the two topological phases $I$ and $II$.
As discussed in ref. \cite{rajak} the revival time scales with the system size. 
At this instant the wave function is peaked around the center of the system and is the result of
a propagating mode across the system with a given velocity and, therefore, scales with the
system size. In the infinite system limit the revival time will diverge and the Majorana mode
decays and is destroyed. A qualitatively different case is illustrated in the last panel of
Fig. \ref{fig2} where a quench from the topological phase $I$ to the quantum critical point at the
origin is considered. In this case the survival probability oscillates indefenitely and periodically
the Majorana mode is revived, even in the infinite size limit. As will be shown next, the
two-dimensional superconductor has some features that are similar, but a richer behavior is found.

\begin{figure}[t]
\includegraphics[width=0.49\columnwidth]{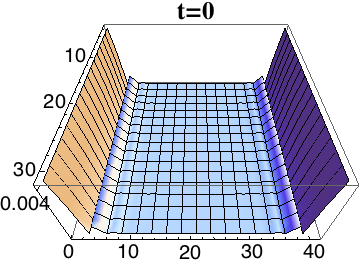}
\includegraphics[width=0.49\columnwidth]{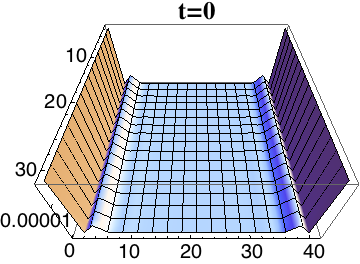}
\includegraphics[width=0.49\columnwidth]{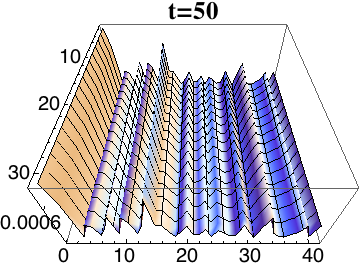}
\includegraphics[width=0.49\columnwidth]{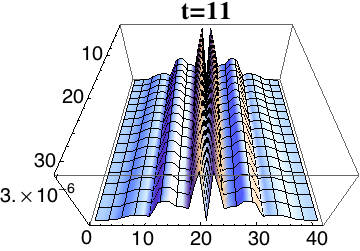}
\includegraphics[width=0.49\columnwidth]{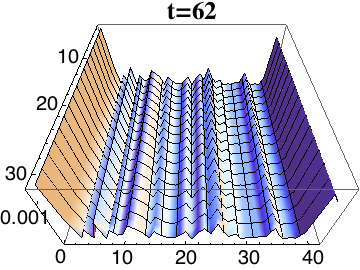}
\includegraphics[width=0.49\columnwidth]{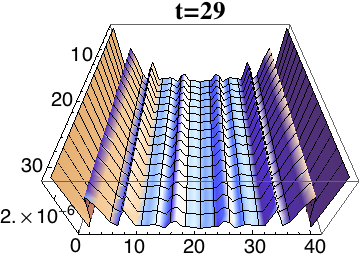}
\caption{\label{fig3}
(Color online) 
Time evolution of real space $|u_{\uparrow}|^2$ for $(M_z=2,\epsilon_F=-5) \rightarrow (M_z=0,\epsilon_F=-5)$, 
$C=1 \rightarrow C=0$ (trivial)
(left column) for $t=0,t=50,t=62$, respectively from top to bottomand
$|u_{\uparrow}|^2$ for $(M_z=3.5,\epsilon_F=0) \rightarrow (M_z=4.5,\epsilon_F=0)$, $C=-2 \rightarrow C=0$ (trivial) (right column)
for $t=0,t=11,t=29$ with strong spin-orbit coupling.
}
\end{figure}

\begin{figure}
\includegraphics[width=0.49\columnwidth]{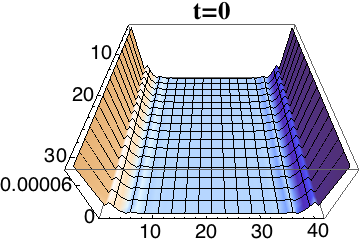}
\includegraphics[width=0.49\columnwidth]{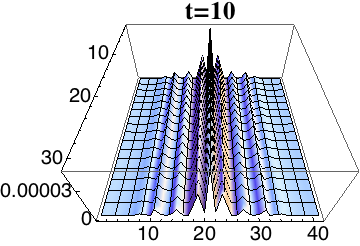}
\includegraphics[width=0.49\columnwidth]{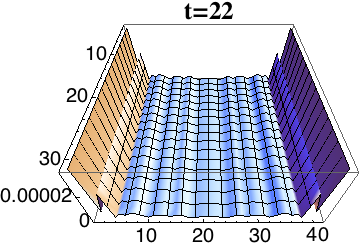}
\caption{\label{fig5}
(Color online) 
Time evolution of real space $|u_{\uparrow}|^2$ for $(Mz=2,\epsilon_F=-1) \rightarrow (M_z=4,\epsilon_F=-1)$, $C=-2 \rightarrow C=-1$ 
for $t=0,t=10,t=22$
with strong spin-orbit coupling.
}
\end{figure}

\subsection{Strong spin-orbit coupling}

\begin{figure}[t]
\includegraphics[width=0.95\columnwidth]{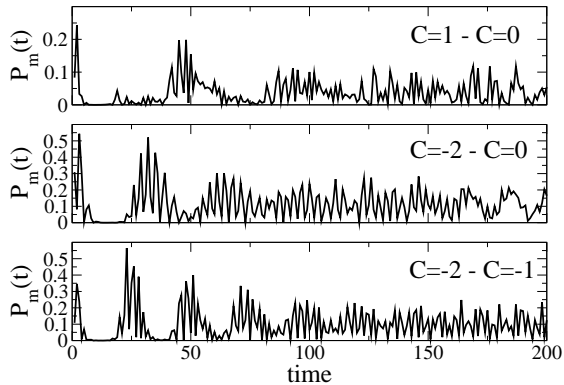}
\caption{\label{fig6}
(Color online) 
Survival probability of the Majorana state of the two-dimensional triplet superconductor
for different transitions across the phase diagram: 
i) transition $(M_z=2,\epsilon_F=-5) \rightarrow (M_z=0,\epsilon_F=-5)$, $C=1 \rightarrow C=0$ (trivial),
ii) transition $(M_z=3.5,\epsilon_F=0) \rightarrow (M_z=4.5,\epsilon_F=0)$, $C=-2 \rightarrow C=0$ (trivial) and
iii) transition $(M_z=2,\epsilon_F=-1) \rightarrow (M_z=4,\epsilon_F=-1)$, $C=-2 \rightarrow C=-1$.
}
\end{figure}

We consider first the case of the two-dimensional superconductor when the spin-orbit coupling is strong which favors that the
superconducting pairing is aligned along $\boldsymbol{s}$. The phase diagram depicted
in Fig. \ref{fig1} applies to this case.

The solutions of the wave functions, written in the form
of a 4-component spinor in real space can be detailed as
\be
\left(\begin{array}{c}
u_n\\
v_n
\end{array}\right)=
\left(\begin{array}{c}
u_n(j_x,j_y,\uparrow) \\
u_n(j_x,j_y,\downarrow) \\
v_n(j_x,j_y,\uparrow) \\
v_n(j_x,j_y,\downarrow) \\
\end{array}\right) 
\label{spinorrs}.\ee
The time evolution of each spinor is given by eq. (\ref{evolve}).
Focusing our attention on the Majorana mode, we present in Figs. (\ref{fig3},\ref{fig5})
the time evolution of the absolute value of the spinor component 
$u_n(j_x,j_y,\uparrow)$, as an example. The other spinor components have a qualitatively similar behavior. 
We consider a system of size $31 \times 41$. A set of characteristic time values are selected.
The initial state shows a mode that is very much peaked at the borders
of the system and that decays fast inside the supercondutor along the transverse
direction. 
As time evolves the peaks move towards the center until they merge at some later time, dependent
of the system transverse size (as for the Kitaev model), as shown in the second panel.
After this time the peaks move from the center, the wave functions become more extended as a mixture
to all the eigenstates becomes more noticeable. Eventually at later times the wave function recovers
a shape that is close to the initial state and there is a revival of the original state. The process
then repeats itself but the same degree of coherence is somewhat lost. In Fig. (\ref{fig3})
the quenches are carried out between a topological phase to a trivial phase ($C=1\rightarrow C=0$) and
to a phase at the border between different phases. In Fig. (\ref{fig5}) the quench is between two
topological phases. In this case the shape of the wave function at the revival time is somewhat
better defined and with a larger overlap to the initial state. A more quantitative description of the
similarity to the initial phase is shown by the survival probability depicted in Fig. (\ref{fig6}).
The three panels show the survival probabilities for the three cases. A behavior similar to the
Kitaev model is shown: after a fast decay of the probability there are revivals of the state at later
times. 

In a way similar to the Kitaev $1d$ case, we may take quenches to the borders in the phase
diagram betwen different phases. At these (quantum critical) points the gap vanishes. It can be shown
that in general the Majorana modes are non-robust but in some cases they are robust.
As an example of a robust Majorana mode a quench $(M_z=1,\epsilon_f=-4) \rightarrow (M_z=1,\epsilon_F=-3)$ 
shows oscillations similar to the Kitaev model (not shown in the figures).

\subsection{Weak spin-orbit coupling}

\begin{figure}[t]
\includegraphics[width=0.49\columnwidth]{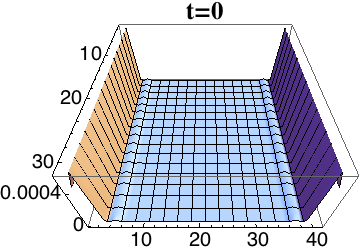}
\includegraphics[width=0.49\columnwidth]{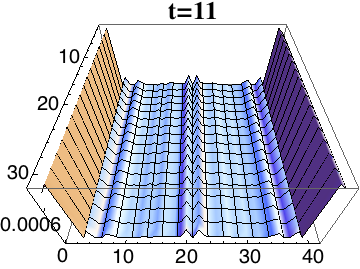}
\includegraphics[width=0.49\columnwidth]{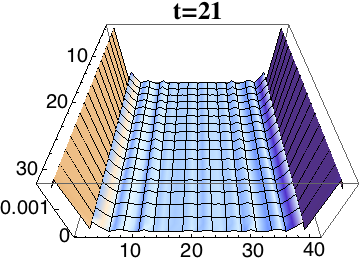}
\caption{\label{fig8}
(Color online) 
Time evolution of real space $|u_{\uparrow}|^2$ for weak spin-orbit coupling 
($\alpha=1,\epsilon_F=-1,d=1,\Delta_s=0$
and $M_z=1.2 \rightarrow M_z=0.5$, $C=2 \rightarrow C=0$ for $t=0,t=11,t=21$.
}
\end{figure}

\begin{figure}
\includegraphics[width=0.95\columnwidth]{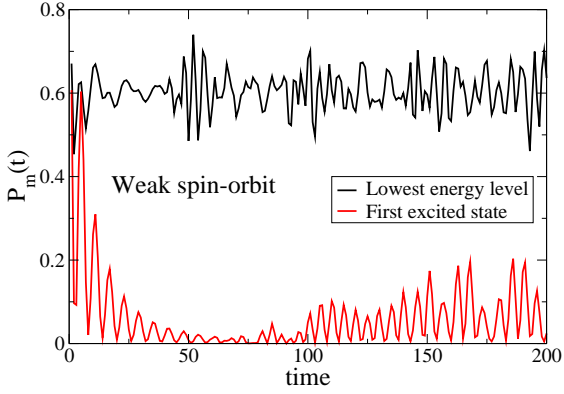}
\caption{\label{fig9}
(Color online) 
Survival probability of the Majorana state of the two-dimensional triplet superconductor
for weak coupling: $\alpha=1,\epsilon_F=-1,\Delta_s=0, d=1$ and initial state $M_z=1.2$
and new parameter set $M_z=0.5$ (corresponds to a transition from $C=2$ to $C=0$).
}
\end{figure}

Consider now the case of weak spin-orbit coupling for which the
pairing is not aligned with the spin-orbit vector. We consider
the case discussed above. The phase diagram for this pairing vector
is the same as for the strong coupling case but the Chern numbers
change signs. In Fig. (\ref{fig8}) we present the time evolution
of the wave function for a case where there is a quench from a
phase with $C=2$ to a trivial phase with $C=0$. The results show that
the Majorana is fairly robust. The time evolution follows the same trends:
after the initial state with a sharp edge state, the shape of the state
is such that ripples appear along the transverse direction that approach the middle
point, scatter each other and propagate back to the border, such as in the
revival process of the strong coupling case. However, throughout this process
the peaks at the border remain. Moreover, at the revival time the wave function
has a shape very close to the initial state. The Majorana state is therefore
quite robust. This is also illustrated in Fig. (\ref{fig9}) where we show the
survival probability of the two lowest levels (the lowest level being the Majorana
fermion). The survival probability of the Majorana fermion is clearly finite,
even though there is a change in topology.

Even though the edge states are fairly localized near the borders, in a finite system
the two edges are coupled. It is therefore convenient to consider a momentum space description
along the longitudinal direction and resolve the modes in $k_x$ space, which allows to solve
larger systems.

\section{Stability of edge states}

We consider a strip geometry of transversal
width $N_y$ and apply  periodic boundary conditions along the longitudinal direction, $x$.
We write
\be
\psi_{k_x,k_y,\sigma} = \frac{1}{\sqrt{N_y}} \sum_{j_y} e^{-i k_y j_y} \psi_{k_x,j_y,\sigma}\,,
\label{operators}
\ee
and rewrite the Hamiltonian matrix in terms of
the operators (\ref{operators})  as
\bea
H = \sum_{k_x} \sum_{j_y}
& & \left(\begin{array}{cccc}
\psi_{k_x,j_y,\uparrow}^{\dagger}  & \psi_{k_x,j_y,\downarrow}^{\dagger} &
\psi_{-k_x,j_y,\uparrow}  & \psi_{-k_x,j_y,\downarrow}
\end{array}\right) \nonumber \\
& & \hat{H}_{k_x,j_y}
\left(\begin{array}{c}
\psi_{k_x,j_y,\uparrow} \\
\psi_{k_x,j_y,\downarrow} \\
\psi_{-k_x,j_y,\uparrow}^{\dagger} \\
\psi_{-k_x,j_y,\downarrow}^{\dagger} \\
\end{array}\right)
\eea

The operator $\hat{H}_{k_x,j_y}$ reads
\be
\hat{H}_{k_x,j_y}=
\left(\begin{array}{cc}
A & B \\
C & D \\
\end{array}\right)
\ee
where $A$ is given by
\be
\left(\begin{array}{cc}
-2 \tilde{t} \cos k_x -M_z-\epsilon_F-\tilde{t} \eta_+ & i\alpha \sin k_x +\frac{\alpha}{2i} \eta_- \\
-i \alpha \sin k_x +\frac{\alpha}{2i} \eta_- & -2 \tilde{t} \cos k_x +M_z -\epsilon_F -\tilde{t} \eta_+ \\
\end{array}\right)
\ee
\be
B=
\left(\begin{array}{cc}
-i d \sin k_x -\frac{d}{2i} \eta_- & 0 \\
0 & -i d \sin k_x +\frac{d}{2i} \eta_- \\
\end{array}\right)
\ee
\be
C=
\left(\begin{array}{cc}
i d \sin k_x -\frac{d}{2i} \eta_- & 0 \\
0 & i d \sin k_x +\frac{d}{2i} \eta_- \\
\end{array}\right)
\ee
and $D$ is given by
\be
\left(\begin{array}{cc}
2 \tilde{t} \cos k_x +M_z+\epsilon_F+\tilde{t} \eta_+ & -i\alpha \sin k_x +\frac{\alpha}{2i} \eta_- \\
i \alpha \sin k_x +\frac{\alpha}{2i} \eta_- & 2 \tilde{t} \cos k_x -M_z +\epsilon_F +\tilde{t} \eta_+ \\
\end{array}\right)
\ee
where $\psi_{j_y}^{\dagger} \eta_{\pm} \psi_{j_y} = \psi_{j_y}^{\dagger} \psi_{j_y+1} \pm \psi_{j_y+1}^{\dagger} \psi_{j_y}$.
The diagonalization of this Hamiltonian involves the solution of a $4 N_y \times 4 N_y$ eigenvalue problem.
The energy states include states in the bulk and states along the edges.

For negative values of the chemical potential, in the topological phases, there is
typically a Majorana mode at $k_x=0$. Other modes appear at other momenta values,
depending on the region in the phase diagram. Since the Hamiltonian factorizes in
momentum space, the overlaps between the single particle states of the initial
Hamiltonian and those of the final Hamiltonian, are restricted to the same momentum
value. Therefore, only states with the same momentum are coupled. We may therefore
follow the time evolution of a given state with given momentum, separately from
other states at different momenta.

\begin{figure}[t]
\includegraphics[width=0.95\columnwidth]{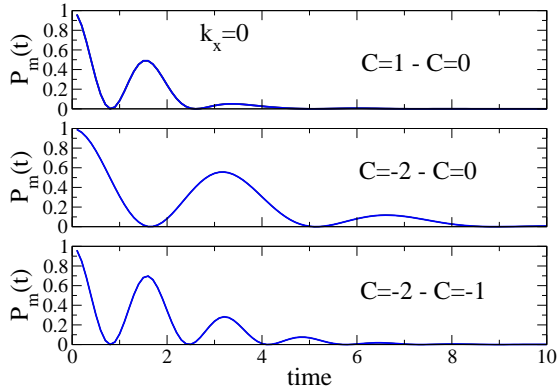}
\caption{\label{fig51}
(Color online) 
Early time survival probability of the Majorana state of the two-dimensional triplet superconductor
for several quenches and different system sizes along the transverse, $y$, direction for $k_x=0$.
The fast decay of the survival probability is independent of system size.  
}
\end{figure}

In Fig. (\ref{fig51}) we consider the same quenches as in Fig. (\ref{fig6}) 
focusing on the survival probability of a Majorana at $k_x=0$. The decay depends
on the parameters chosen but is independent of the transvere direction system size,
for short times. The results for system sizes from $N_y=20$ to $N_y=200$ are superimposed.
The decay of the Majorana mode is therefore independent of system size, as expected of
a localized state. Note however that $N_y=20$ is a rather small size.

\begin{figure}[t]
\includegraphics[width=0.65\columnwidth]{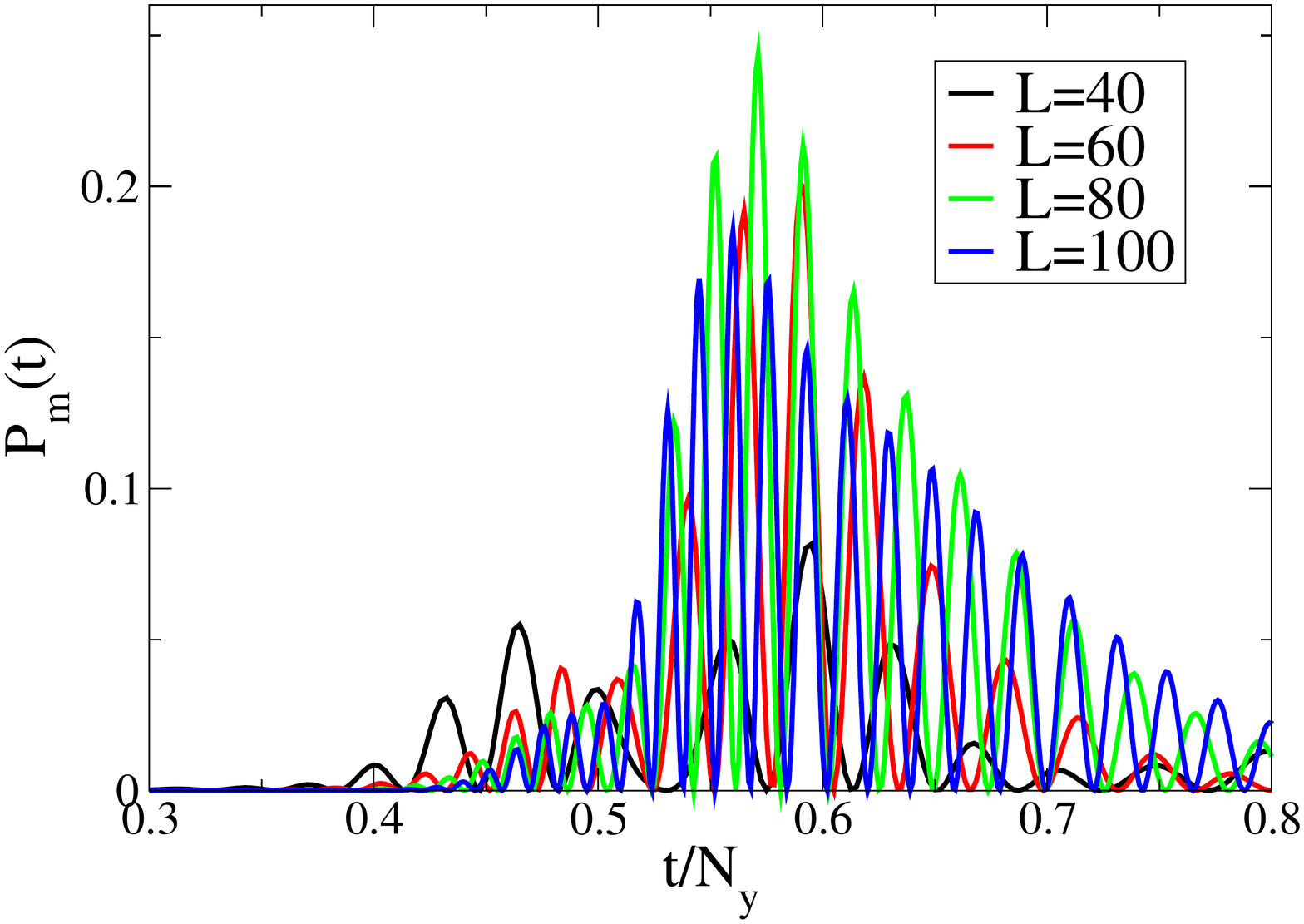}
\includegraphics[width=0.65\columnwidth]{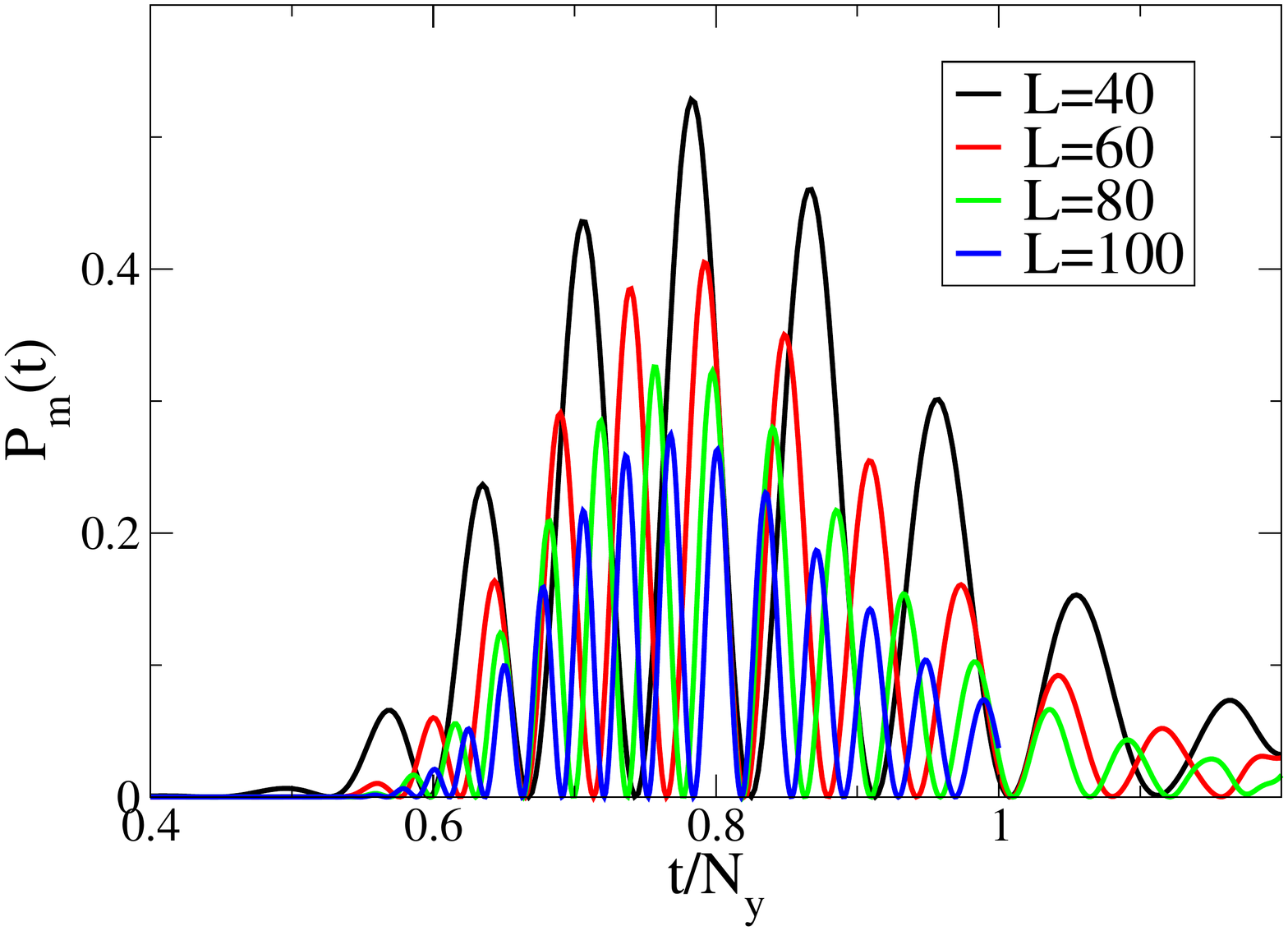}
\includegraphics[width=0.65\columnwidth]{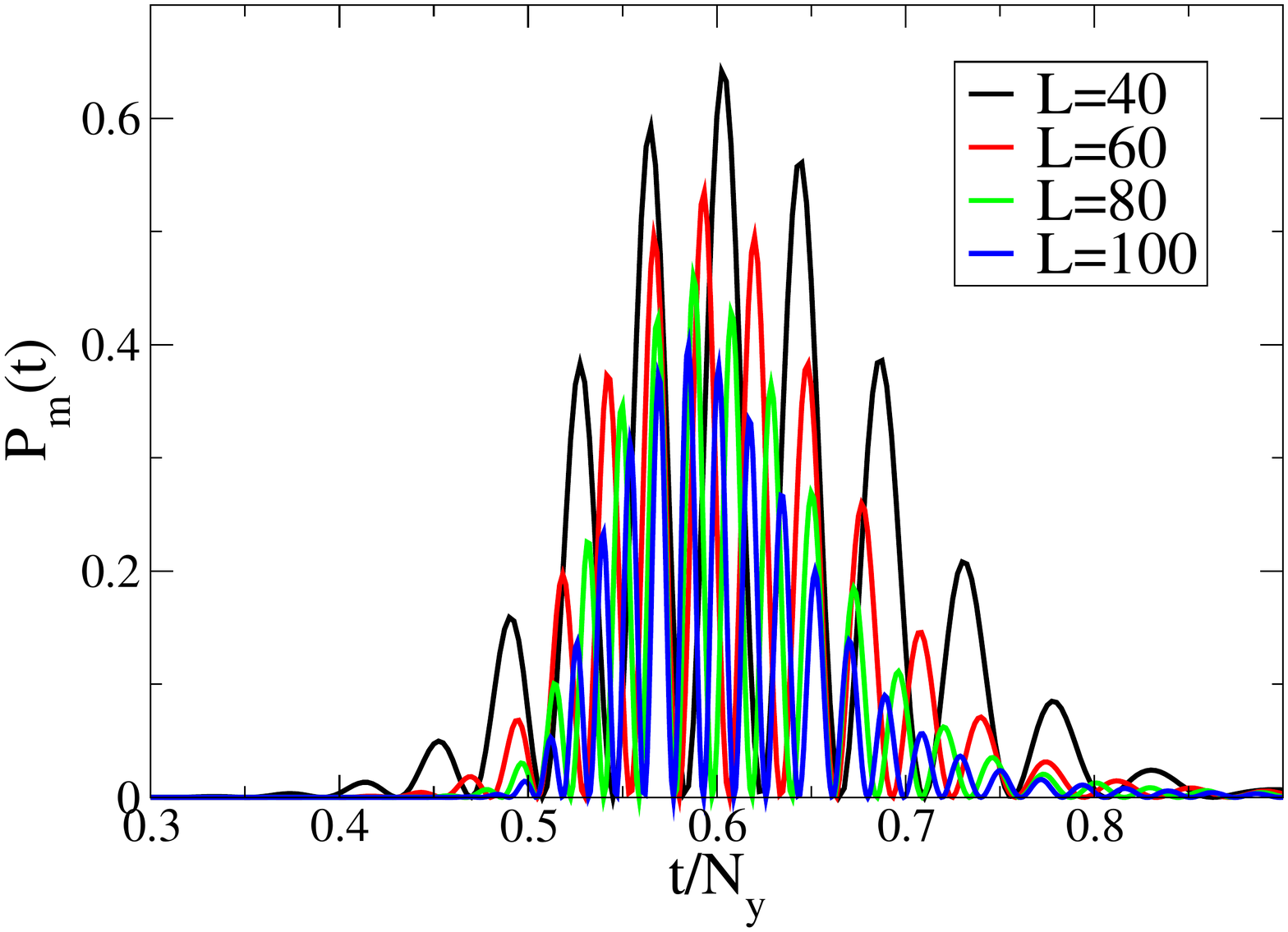}
\caption{\label{fig52}
(Color online) 
Scaling of the survival probability of the Majorana state of the two-dimensional triplet superconductor
for several quenches and different system sizes along the transverse, $y$, direction for $k_x=0$.
The first revival time scales with the system transverse direction, due to the propagation
of the mode to the center of the system.
}
\end{figure}

On the other hand, as mentioned above, the instant of the merging of the peaks of the wave function
is size dependent. In Fig. (\ref{fig52}) we present the scaling of the collision time with system
size ($t/N_y$) for various system sizes, from $N_y=20$ until $N_y=120$. The approximate
scaling is apparent, confirming the picture observed in the Kitaev, $1d$, problem.
In the following the results will be presented for a system size $N_y=100$.

We reanalize some results obtained above, considering larger system sizes.
Moreover,
the results are resolved in momentum space which provides more information on the details
of the overlaps between the eigenstates of the two Hamiltonians.

\begin{figure}[t]
\includegraphics[width=0.95\columnwidth]{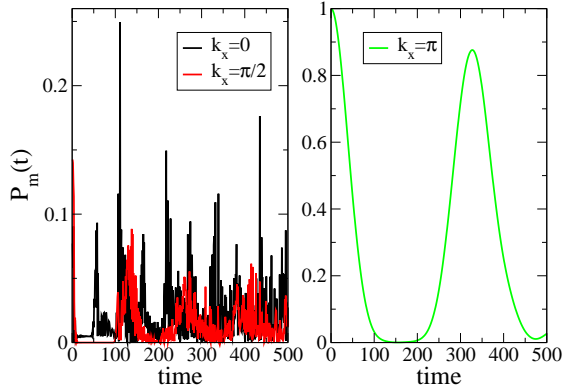}
\caption{\label{fig11}
(Color online) 
Survival probability of the Majorana state of the two-dimensional triplet superconductor
for a quench $(M_z=2,\epsilon_F=-3) \rightarrow (M_z=0,\epsilon_F=-3)$, $C=1$ to $C=0$ ($Z_2$ phase).
}
\end{figure}

Considering a quench from a $C=1$ phase to a trivial phase leads to the same 
fast decay of the Majorana mode and a mode revival that scales to infinity as the system size
grows. However, as shown in Fig. (\ref{fig11}), taking a quench to a $Z_2$ phase (even though
the Chern number vanishes, there are edge states at the same momentum value $k_x=0$), 
we can see that, even though the decay of the mode is sharp, and the survival probability
is very small it is finite. A similar result is obtained performing a quench from $C=-2$ to a $Z_2$
phase, where a very small probability arises. 
In Fig. (\ref{fig11}) we also show results for the lowest energy states for two other
momenta values $k_x=\pi/2,\pi$. For these parameters these states have finite energy
and are not Majorana zero modes. The mode at $k_x=\pi/2$ shows similar decay/revival behavior
but the mode at $k_x=\pi$ shows oscillatory behavior.
Interesting behavior is also found for the mode at momentum $k_x=\pi$ when a transition
between an initial state with a Majorana fermion at $k_x=0$ and a final state with a Majorana
fermion at momentum $k_x=\pi$ occurs. This happens, for instance, in the quenches
$(M_z=3,\epsilon_F=-4) \rightarrow (M_z=3,\epsilon_F=-2)$, for which $C=1\rightarrow C=-1$, or
$(M_z=2,\epsilon_F=-3) \rightarrow (M_z=2,\epsilon_F=-1)$, for which $C=1\rightarrow C=-2$.
In these cases, while the $k_x=0$ mode decays, the finite energy state at momentum $k_x=\pi$ 
has a survival probability that is unity. The coupling to the Majorana state of the final state
Hamiltonian is therefore unity. 

A study of the projections between the two sets of
eigenstates, 
$| \langle \psi_m(\xi)|\psi_n(\xi^{\prime})\rangle|^2$,
shows that, in general, the overlaps are rather small, except for a few selected
states. As expected, in quenches with strong decay all overlaps are quite small,
particularly with the low energy modes, for each momentum value. The opposite cases of
finite survival probabilities are associated with larger overlaps, typically with a few
states. Considering the $k_x=0$ state, this robustness is usually associated with a
large overlap to a final Hamiltonian eigenstate of small energy. However, in some
cases there is a large overlap to states at finite energies, but whose
wave functions are somewhat similar to the edge states (most likely they are antibound states
between the two edges of the system and lie near gaps that appear in the spectrum at finite energies).

\begin{figure}[t]
\includegraphics[width=0.95\columnwidth]{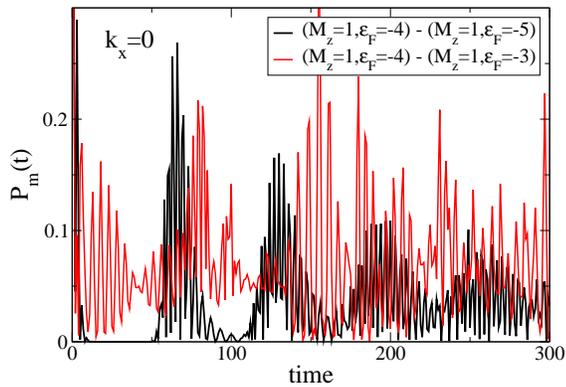}
\caption{\label{fig17}
(Color online) 
Survival probability of the Majorana state of the two-dimensional triplet superconductor
for a quench $C=1$ to quantum critical points with $C=0$ (trivial and with edge states).
$(M_z=1,\epsilon_F=-4) \rightarrow (M_z=1,\epsilon_F=-5)$,  and $(M_z=1,\epsilon_F=-4) \rightarrow 
(M_z=1,\epsilon_F=-3)$.
}
\end{figure}

\begin{figure}[t]
\includegraphics[width=0.65\columnwidth]{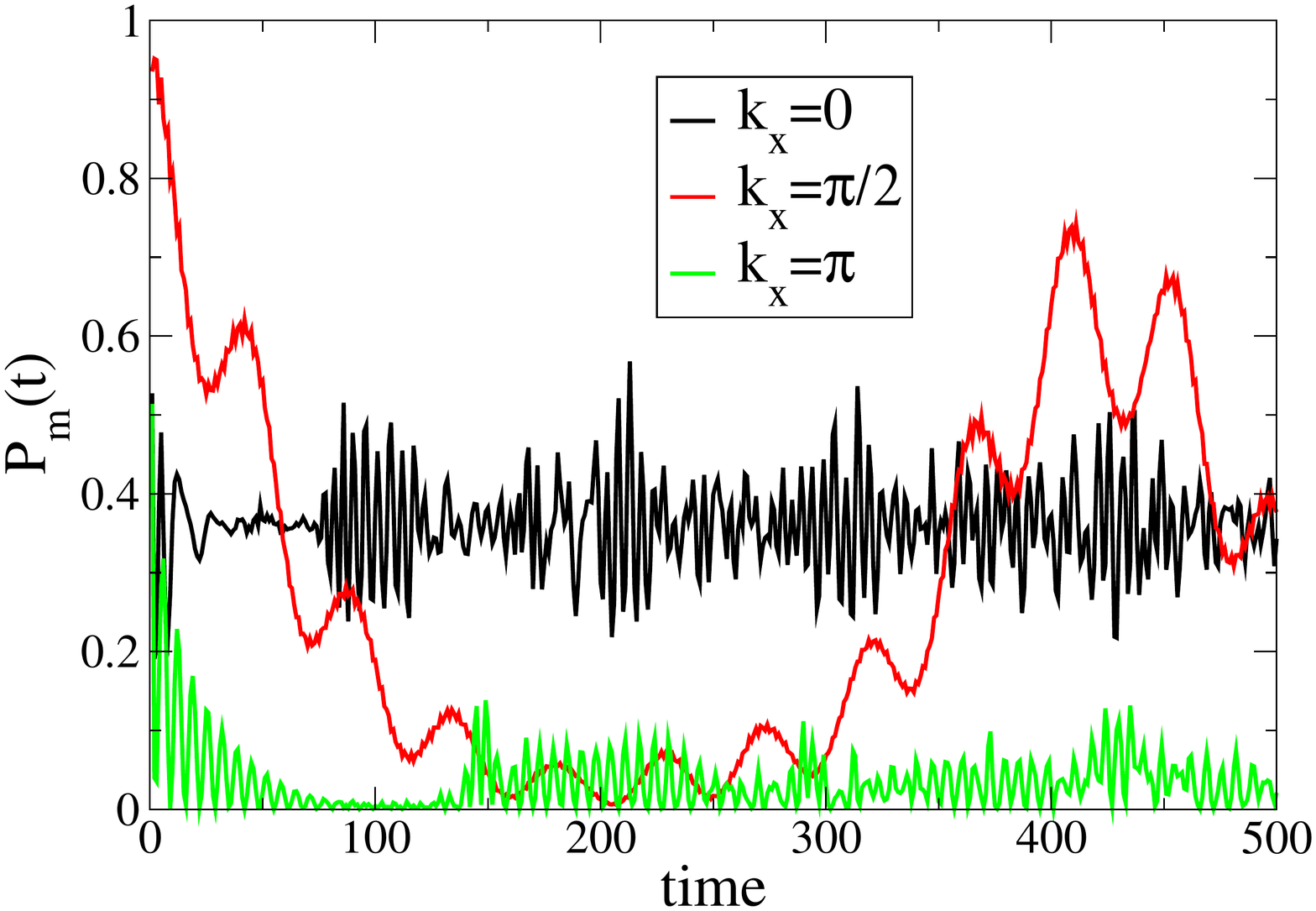}
\includegraphics[width=0.65\columnwidth]{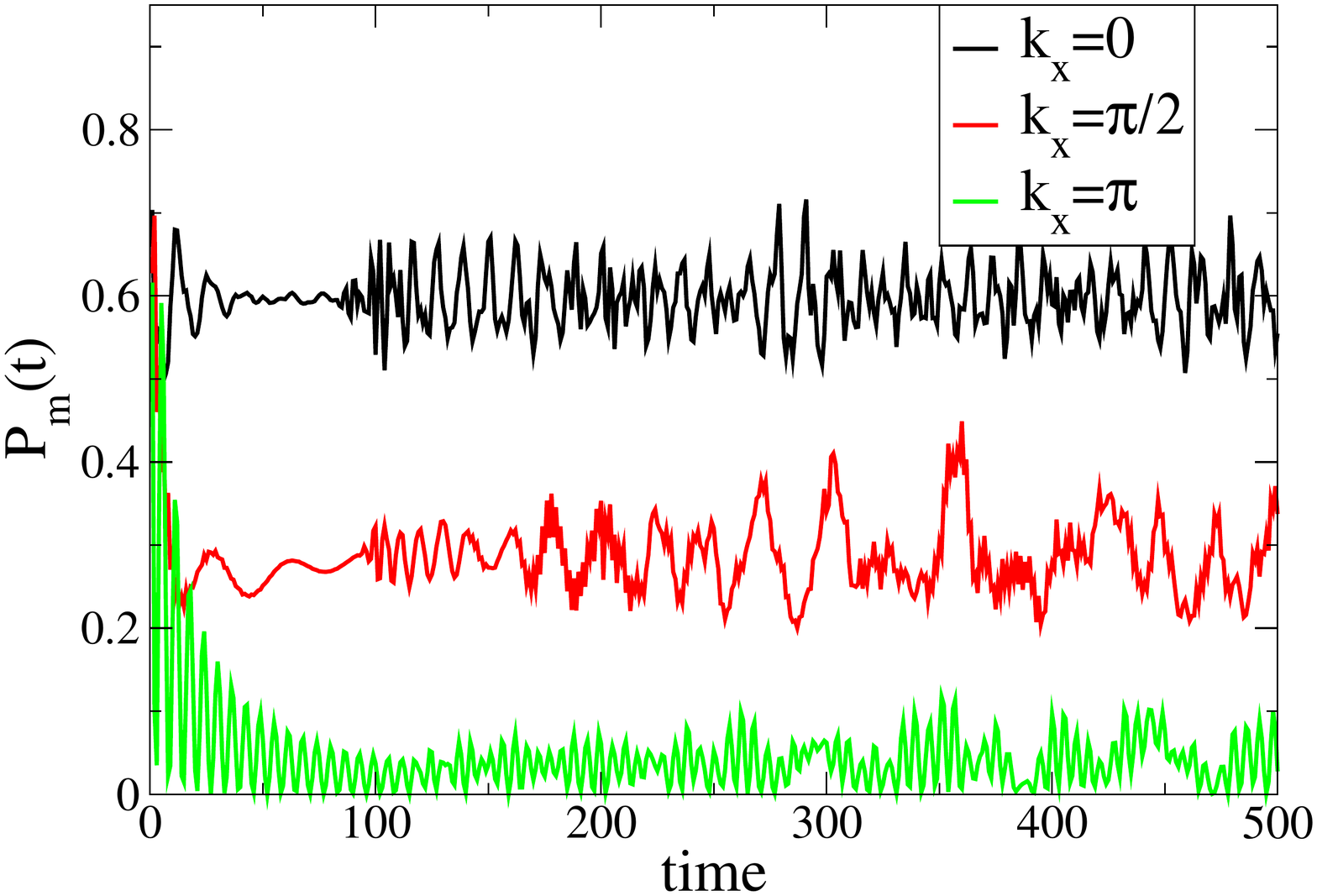}
\includegraphics[width=0.65\columnwidth]{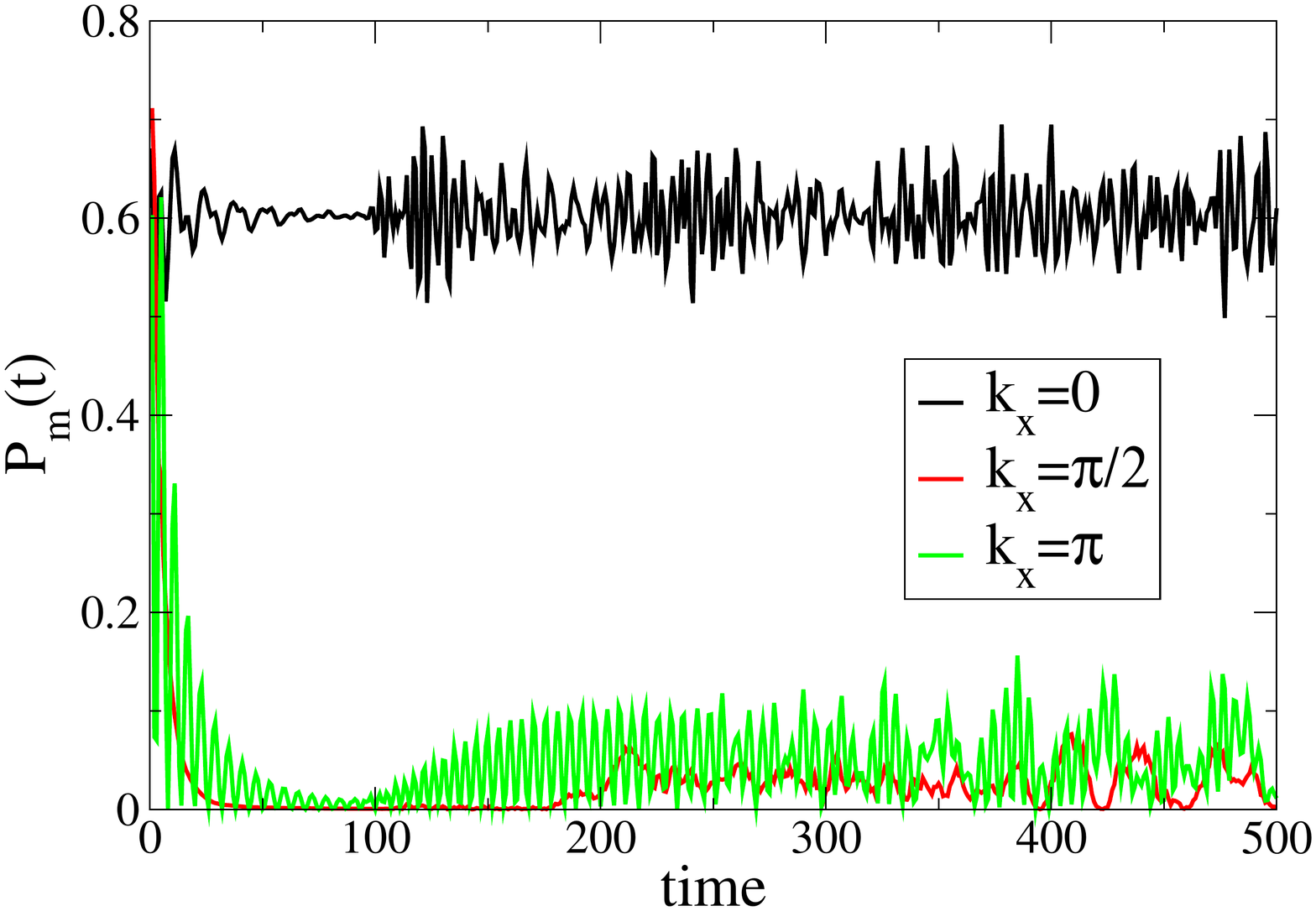}
\caption{\label{fig18}
(Color online) 
Survival probability of the Majorana state of the two-dimensional triplet superconductor
for the case of weak spin-orbit coupling with
$\epsilon_F=-1,d=1,\Delta_s=0$ and
a) $(\alpha=3,M_z=1.2) \rightarrow (\alpha=2,M_z=0.5)$,
b) $(\alpha=1.5,M_z=1.2) \rightarrow (\alpha=1.5,M_z=0.5)$ and
c) $(\alpha=1,M_z=1.2) \rightarrow (\alpha=1,M_z=0.5)$
corresponding to $C=2 \rightarrow C=0$ for all cases.
}
\end{figure}

To stress the relevance of the presence of edge states in the same momentum subspace we consider
quenches from a topological phase to the border of other phases (quantum critical points).
Specifically, we consider as initial state $(M_z=1,\epsilon_F=-4)$ located in a phase with
$C=1$. Two quenches are considered, one to the frontier to a trivial phase, ($C=0$ and no edge
states) with $(M_z=1,\epsilon_F=-5)$ and another to the frontier to a non-trivial phase,
($C=0$, but with edge state at $k_x=0$). The results are presented in Fig. (\ref{fig17}). The
survival probability decays in the case of the trivial phase but remains finite in the other case.

As suggested by the results for the smaller system sizes (obtained above in the real-space description), 
in the case of weak spin-orbit coupling
the Majorana modes are considerably more robust. This is illustrated in Fig. (\ref{fig18}) where results
for various quenches are presented. We present results for the lowest energy states
for the same set of momenta $(k_x=0,k_x=\pi/2,k_x=\pi)$. In the various cases there is a zero
energy mode at $k_x=0$ in the initial state as well as a zero energy mode at $k_x=\pi$ (two zero modes
since $C=2$). In the final states there are two zero energy modes at $k_x=0$ but not at either
$k_x=\pi/2$ or $k_x=\pi$. The results are again consistent. The survival probability is finite
for the $k_x=0$ case. This Majorana mode is robust in these quenches. The Majorana mode at
$k_x=\pi$ decays.

\section{Evolution of Chern numbers}

The topology of each phase may be characterized by the Chern number.
As the system evolves in time, the wave functions change. Solving for the evolution
of the wave functions we may calculate the Chern number as a function of time and
determine how the topology changes as well.

\begin{figure}[t]
\includegraphics[width=0.95\columnwidth]{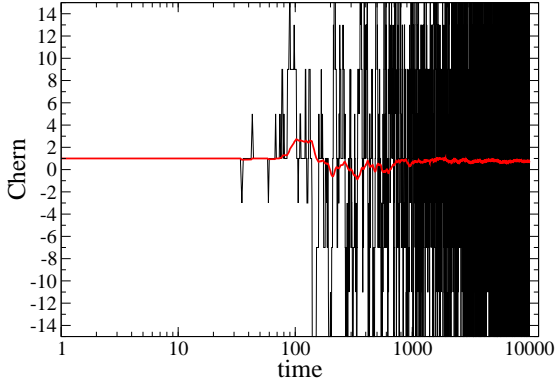}
\caption{\label{fig27}
(Color online) 
Chern numbers for strong spin-orbit coupling for
$(M_z=2,\epsilon_F=-5) \rightarrow (M_z=0,\epsilon_F=-5)$, corresponding to $C=1 \rightarrow C=0$.
}
\end{figure}

\begin{figure}[t]
\includegraphics[width=0.95\columnwidth]{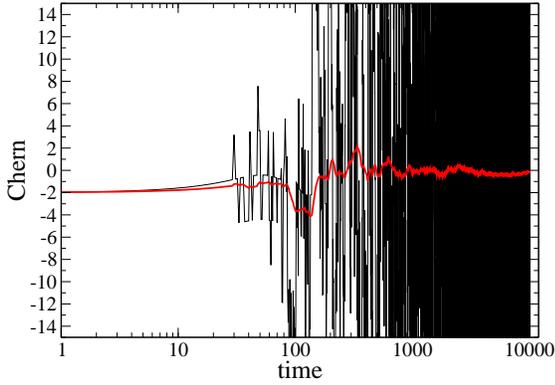}
\caption{\label{fig28}
(Color online)
Chern numbers for strong spin-orbit coupling for
$(M_z=2,\epsilon_F=-1) \rightarrow (M_z=0,\epsilon_F=-1)$, corresponding to $C=-2 \rightarrow C=0$.
}
\end{figure}

It is convenient to calculate the Chern number by computing the
flux of the Berry curvature over plaquetes in the Brillouin zone \cite{Fukui}.
Discretizing the Brillouin zone as $k_\mu = 2\pi j/N$, with $j=1,...,N$, and $\mu=x,y$,
a new variable, $T_{\mu}(\vk)$, for the link $\delta k_\mu$ (with $|\delta k_\mu|= 2\pi/N$)
oriented along the  $\mu$ direction from the point $\vk$ may be defined as
\be
T_{\mu}(\vk) = \frac{\langle \psi_n(\vk)|\psi_n(\vk +\delta k_\mu)\rangle}{|\langle \psi_n(\vk)|\psi_n(\vk +\delta k_\mu)\rangle |}\,,
\ee
and the lattice field strength may be defined as
\be
F_{xy}(\vk) = \ln \left( T_x(\vk) T_y(\vk +\delta k_x) T_x(\vk +\delta k_y)^{-1} T_y(\vk )^{-1} \right)\,.
\ee
$F_{xy}(\vk)$  is restricted to the interval
$-\pi < -i F_{xy}(\vk) \leq \pi$ and the gauge invariant expression for the Chern number is
\be
C_n=\frac{1}{2\pi i} \sum_\vk F_{xy}(\vk)\,.
\label{fukui}
\ee
The calculations of the Chern number of each band $n$ are performed in this way in this work.

In Figs. (\ref{fig27},\ref{fig28}) we present results for the Chern numbers for two quenches.
After the quench the Chern number remains invariant at the initial state value.
Beyond a certain time the Chern number starts to oscillate and these oscillations
become increasingly large. Some sense may be achieved by calculating the time average
of the Chern numbers. It is seen that this time average tends to approach, for long times,
the value corresponding to the Chern value of the final state. However, the evolution is
rather slow. In the first panel that corresponds to a transition from a topological phase
with $C=1$ to a trivial phase with $C=0$, even though the average Chern number is decreasing,
even after 10000 time steps it is still quite far from the asymptotic value.
In the second panel the convergence is faster from a $C=-2$ to $C=0$. Other quenches have
been considered, inclusively between two topological phases but the convergence is very
slow and is not conclusive if it fully occurs. Also, the values taken by the Chern number
at a given time can be quite large.

\begin{figure}[t]
\includegraphics[width=0.95\columnwidth]{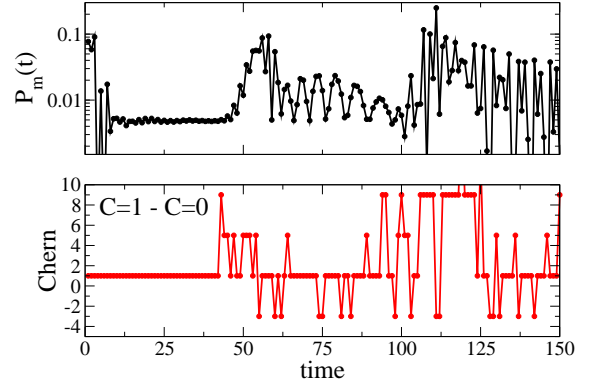}
\caption{\label{figP}
(Color online) 
Comparison of time evolution of Chern number and survival probability.
The Chern number remains stable at the initial state value until
the Majorana mode reaches the middle point of the system. Beyond this
instant the Chern number fluctuates.
}
\end{figure}

In Fig. (\ref{figP}) it is shown that the Chern number remains locked to the initial
state value until the Majorana mode reaches the center point of the system. Beyond that instant
the Chern number starts to fluctuate.

\section{Slow quenches}

Let us consider now a slow transformation of the Hamiltonian
parameters and consider a change that leads the system across
a phase transition. For simplicity we may consider a transformation where only one parameter
changes as time increases as
\be
\xi(t)=\xi(t_0)+r (t-t_0)
\ee
Here $t_0$ is the initial time and $r=(\xi(t_f)-\xi(t_0))/(t_f-t_0)$ is the rate of change
and $t_f$ is the final time. The time evolution of any state is given as before by
\be
|\psi_m(\xi(t))\rangle = U(t,t_0) |\psi_m(\xi)\rangle
\label{evolve2}
\ee
where $U(t,t_0)$ is the time evolution operator. Unlike in the case of the instantaneous
quench considered above, as time goes by the Hamiltonian changes and the eigenstates
change continuosly with time.
We may split the time evolution operator as a path integral
\be
U(t_f,t_0) = U(t_f,t_{N-1}) U(t_{N-1},t_{N-2}) \cdots U(t_2,t_1) U(t_1,t_0)
\ee
dividing the time evolution in $N$ discrete steps. The time evolution of any state
may then be obtained at a set of discrete times, inserting complete sets of eigenstates
of the Hamiltonian at each discrete time. If the number of steps is large enough,
we may with good approximation write
\be
U(t_{i+1},t_i)=e^{-i H(\xi(\bar{t})) \Delta t}
\ee
where $\Delta t=(t_f-t_0)/N$ and $\bar{t}$ is an appropriate time in the interval
$\{t_i,t_{i+1}\}$. For convenience we may take $\bar{t}=t_{i+1}$.
In a way similar to the case of the instantaneous quench, we calculate the overlap
of the time evolved state with the initial state. In particular, we will focus 
attention on the evolution of the single-particle Majorana bound states as the
topology changes across a phase transition.

The overlap amplitude at a given time $t=t_{i}$ can be obtained as
\be
A(t_{i})=\sum_{n_i} \langle \psi_0(\xi(t_0))| \psi_{n_i}(\xi(t_i))\rangle e^{-i E_{n_i}(\xi(t_i)) \Delta t}
A_{n_i}^i
\ee
where
\bea
A_{n_i}^i = \sum_{n_{i-1}} &&  \langle \psi_{n_i}(\xi(t_i))|\psi_{n_{i-1}}(\xi(t_{i-1}))\rangle 
e^{-i E_{n_{i-1}}(\xi(t_{i-1})) \Delta t} \nonumber \\
&& A_{n_{i-1}}^{i-1}
\eea
with
\be
A_{n_1}^1 = \langle \psi_{n_1}(\xi(t_1)) | \psi_0(\xi(t_0)) \rangle
\ee
We will consider the probability defined as
\be
P(t)=|A(t)|^2
\ee

\begin{figure}[t]
\includegraphics[width=0.95\columnwidth]{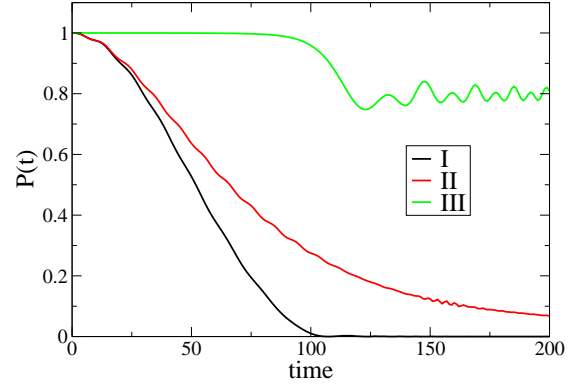}
\caption{\label{fig60}
(Color online)
Time evolution of $P(t)$ for three different cuts in the phase diagram, I, II and III.
corresponding to transitions $C=1\rightarrow C=0$ (trivial phase), $C=1 \rightarrow C=0$
($Z_2$ phase) and $C=-1 \rightarrow C=-2$, respectively. In all cases the transition occurs
at half the time. 
}
\end{figure}

\subsection{Survival probability}

We consider three quantum phase transitions.
We take three cuts represented in Fig. (\ref{fig1}) at constant chemical potential and vary the magnetization.
The first cut, $I$, is obtained keeping $\epsilon_F=-5$ and varying the magnetization from $M_z=2 \rightarrow M_z=0$, 
corresponding to a transition $C=1\rightarrow C=0$ (trivial phase). In the second cut, $II$,
$\epsilon_F=-3$ and the magnetization has the same variation, corresponding to  $C=1 \rightarrow C=0$
($Z_2$ phase). Finally, in the third cut, $III$, $\epsilon_F=-1$ the magnetization varies $M_z=3.5 \rightarrow M_z=2.5$,
across a transition between two topological phases as $C=-1 \rightarrow C=-2$. 
In all cases the transition occurs at half the time interval.

For a given change of the magnetization across the topological transition, the rate is determined by the
time interval. The (discrete) path integral approach used implies a discretization of the time interval.
We have confirmed that the errors introduced by the discretization are very small. Keeping the rate fixed
at a value of the order of $r=0.02$ and changing the number of time steps from $100$ to $1000$ (corresponding
from $\Delta t=1$ to $\Delta t=0.1$) the difference in the results is negligible. 

In Fig. \ref{fig60} the time evolution of $P(t)$ is shown for the three cuts $I,II,III$. These results
were obtained taking a number of time steps $N=200$ and $t_f=200$ which implies $\Delta t=1$ and a rate of $r=2/200$ for
cuts $I,II$ and a rate of $r=1/200$ for cut $III$.
The behavior of the Majorana mode follows similar trends to the abrupt quenches considered in previous
sections. In the case of the transition to the topologically trivial phase ($I$) the overlap tends to zero after the
quantum critical point (half the time interval). Unlike the case of the abrupt quench, there is no revival of the
Majorana state (the system considered here has a size $N_y=100$ and we take $N_x=201$). Since the evolution
closely follows the slow evolution of the Hamiltonian parameters, the state is not recovered after the
transition. In the case of the quench to the $Z_2$ phase, the overlap does not vanish at the final time $t_f$.
In the case of quench $III$ between two topological phases, we see that, until the transition occurs, the
overlap is close to unity; after the transition it decreases, but remains finite. We note however, that
as the slow rate decreases, the overlap decreases as well and in the infinite time limit it seems to converge
to zero.

\begin{figure}[t]
\includegraphics[width=0.49\columnwidth]{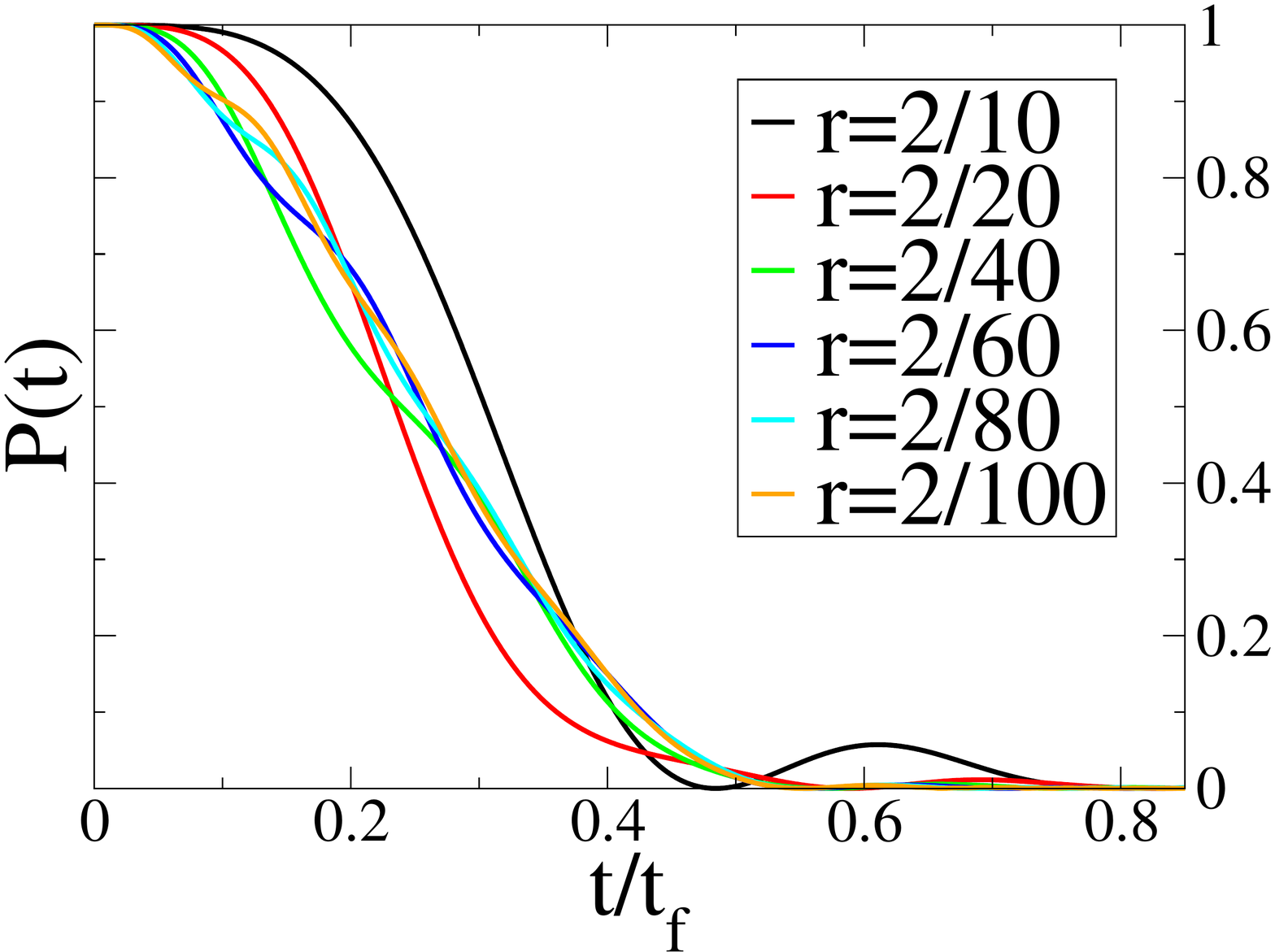}
\includegraphics[width=0.49\columnwidth]{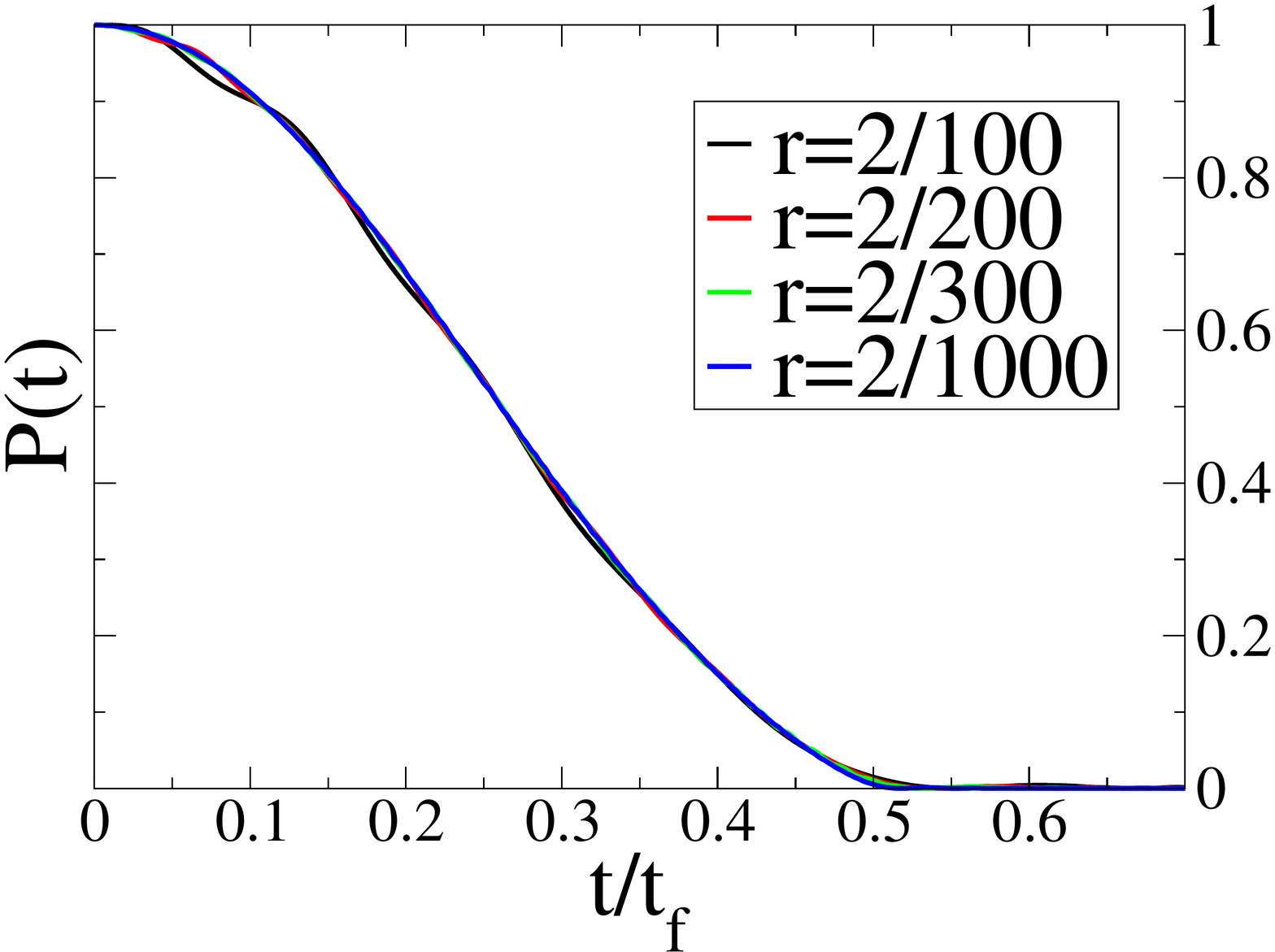}
\includegraphics[width=0.49\columnwidth]{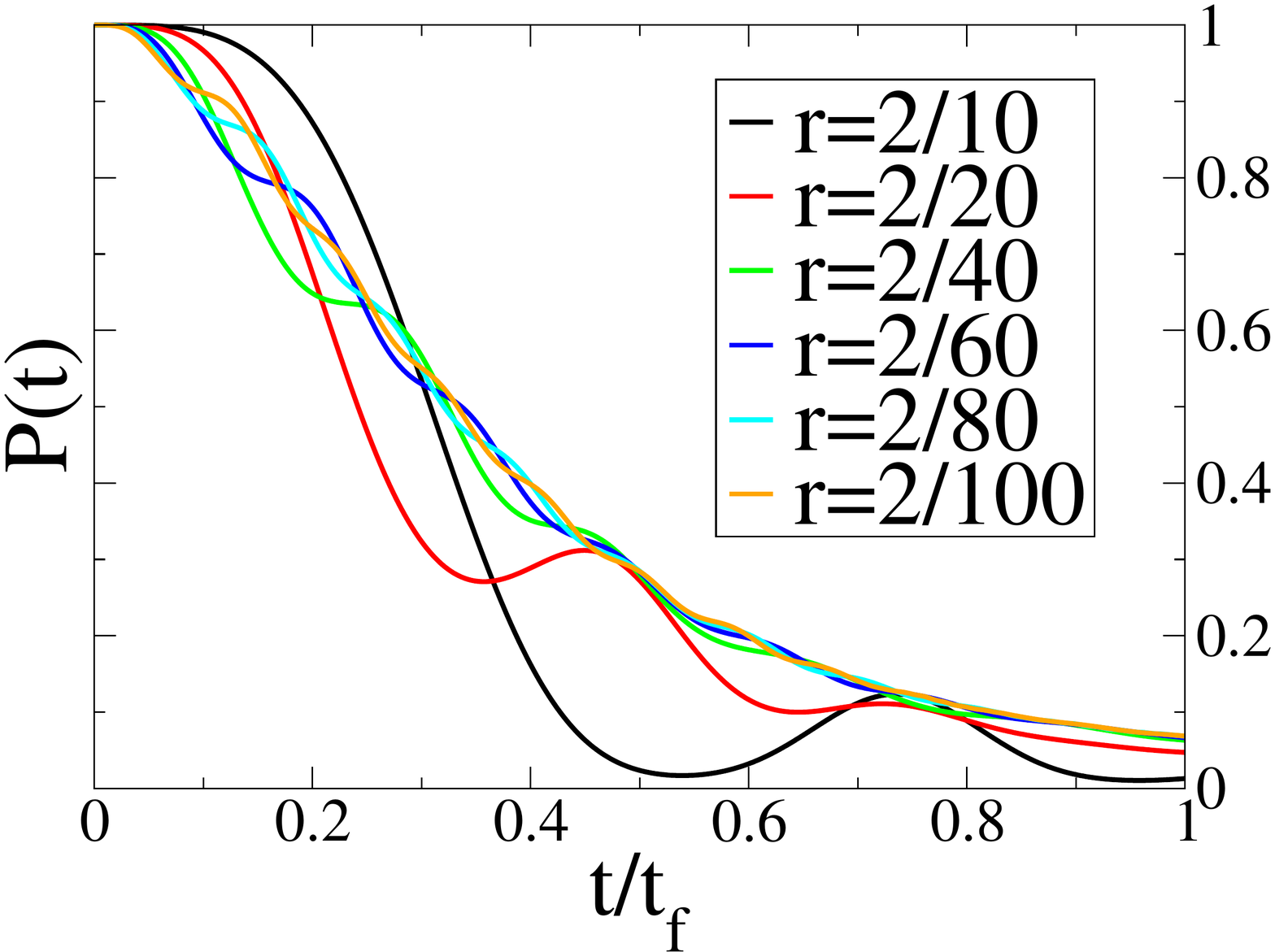}
\includegraphics[width=0.49\columnwidth]{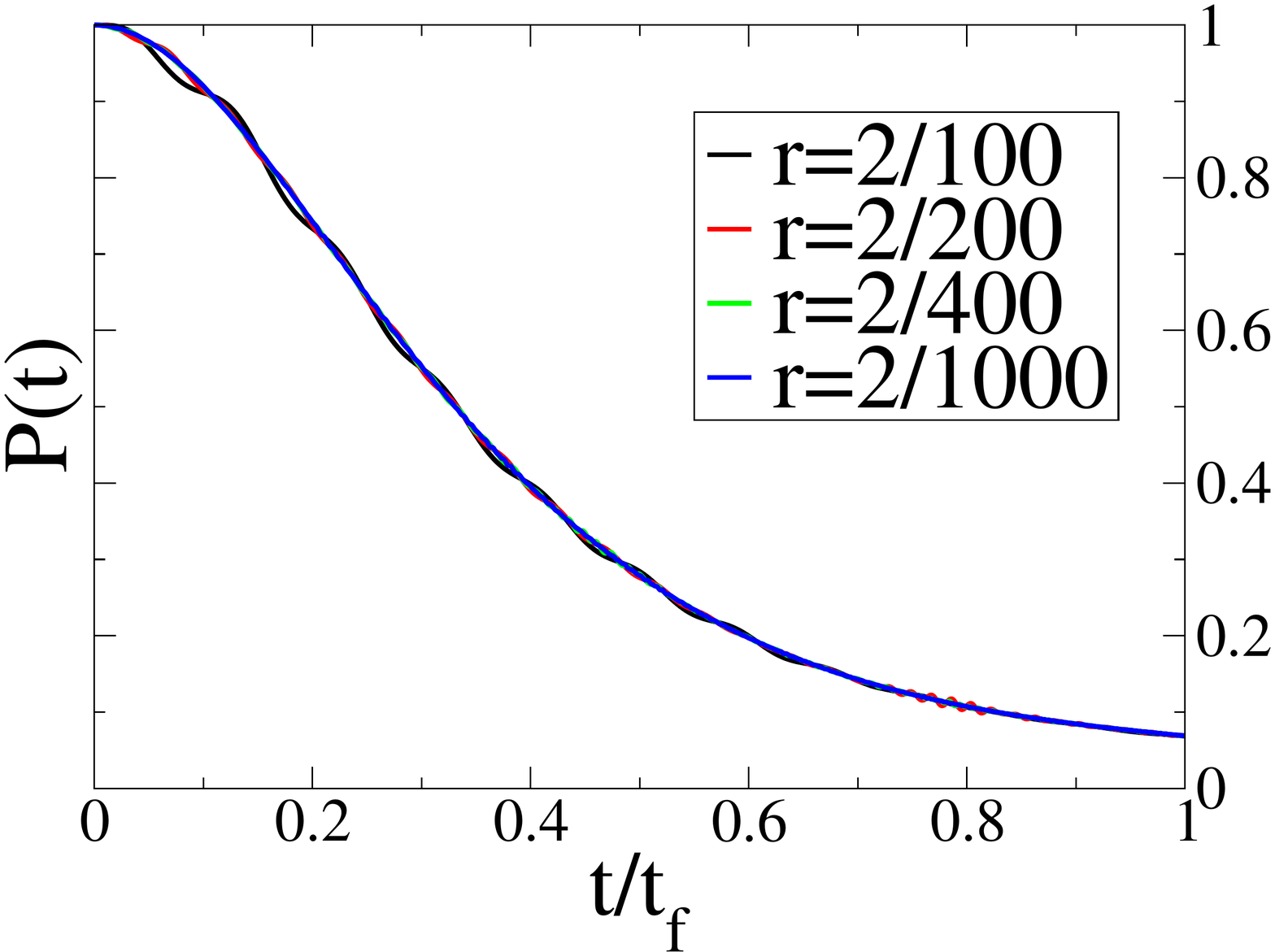}
\caption{\label{fig61}
(Color online)
Time evolution of $P(t)$ for cuts I (upper row) and II (lower row), for different rates.
}
\end{figure}

\begin{figure}[t]
\includegraphics[width=0.49\columnwidth]{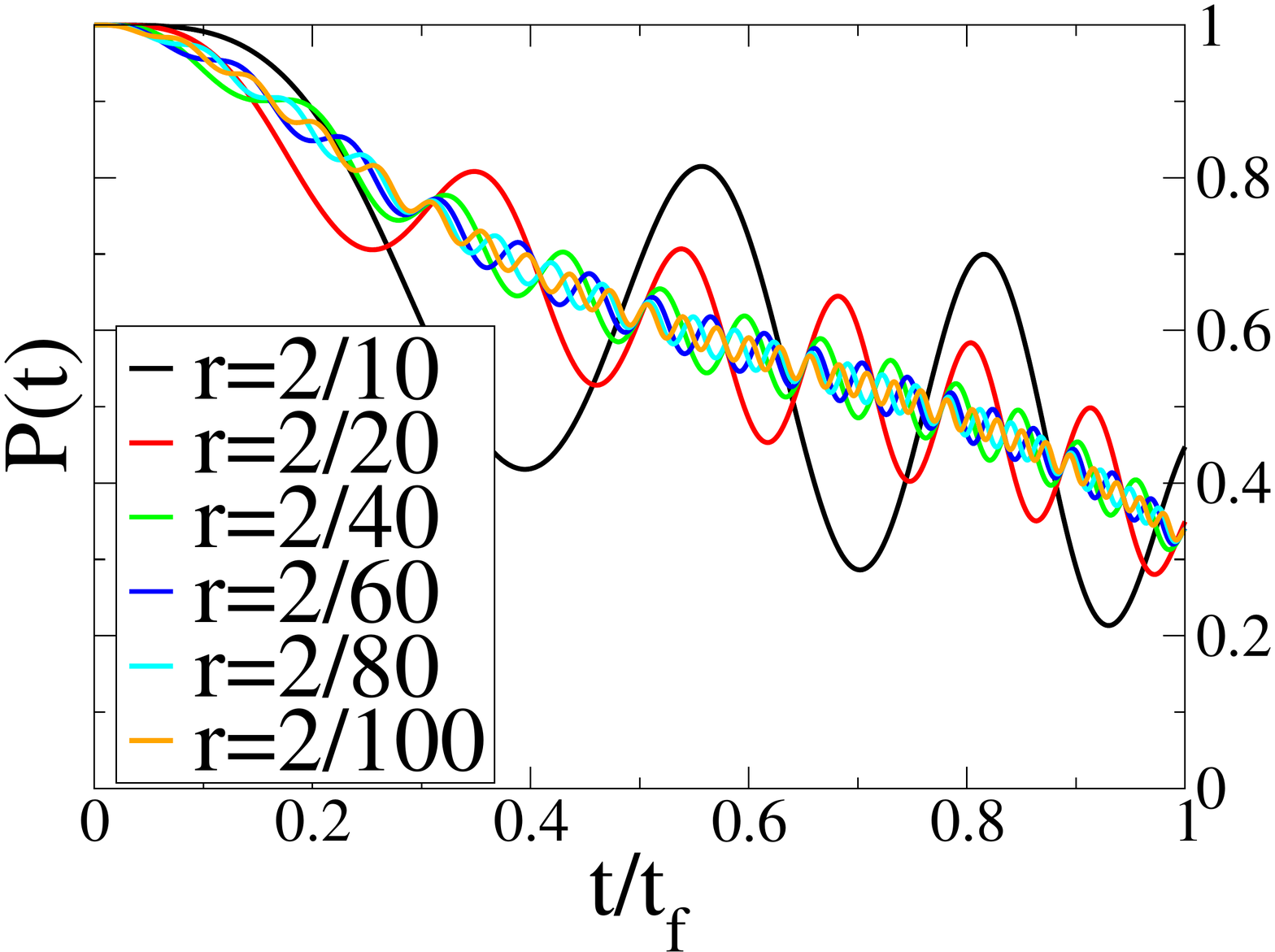}
\includegraphics[width=0.49\columnwidth]{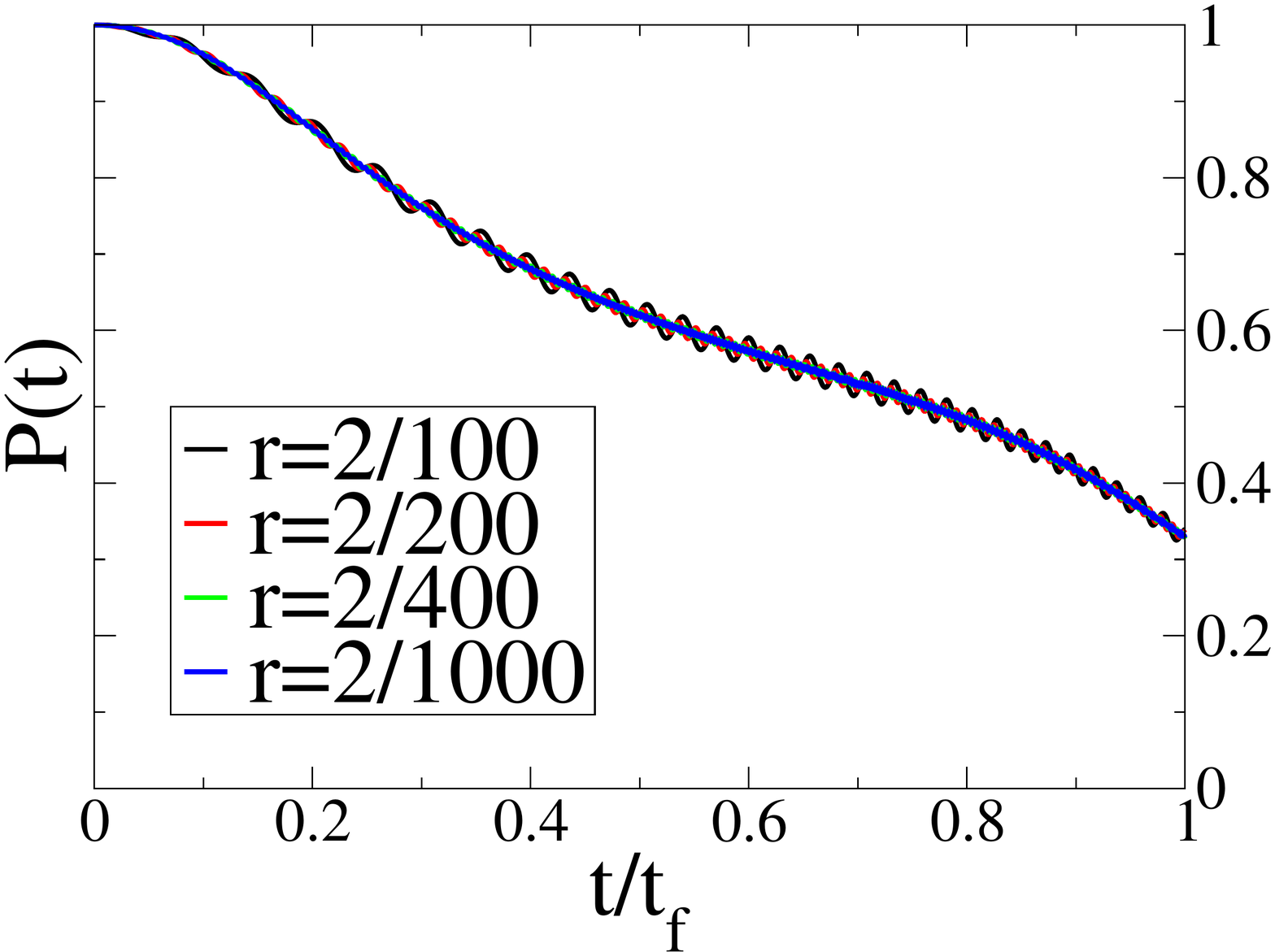}
\includegraphics[width=0.49\columnwidth]{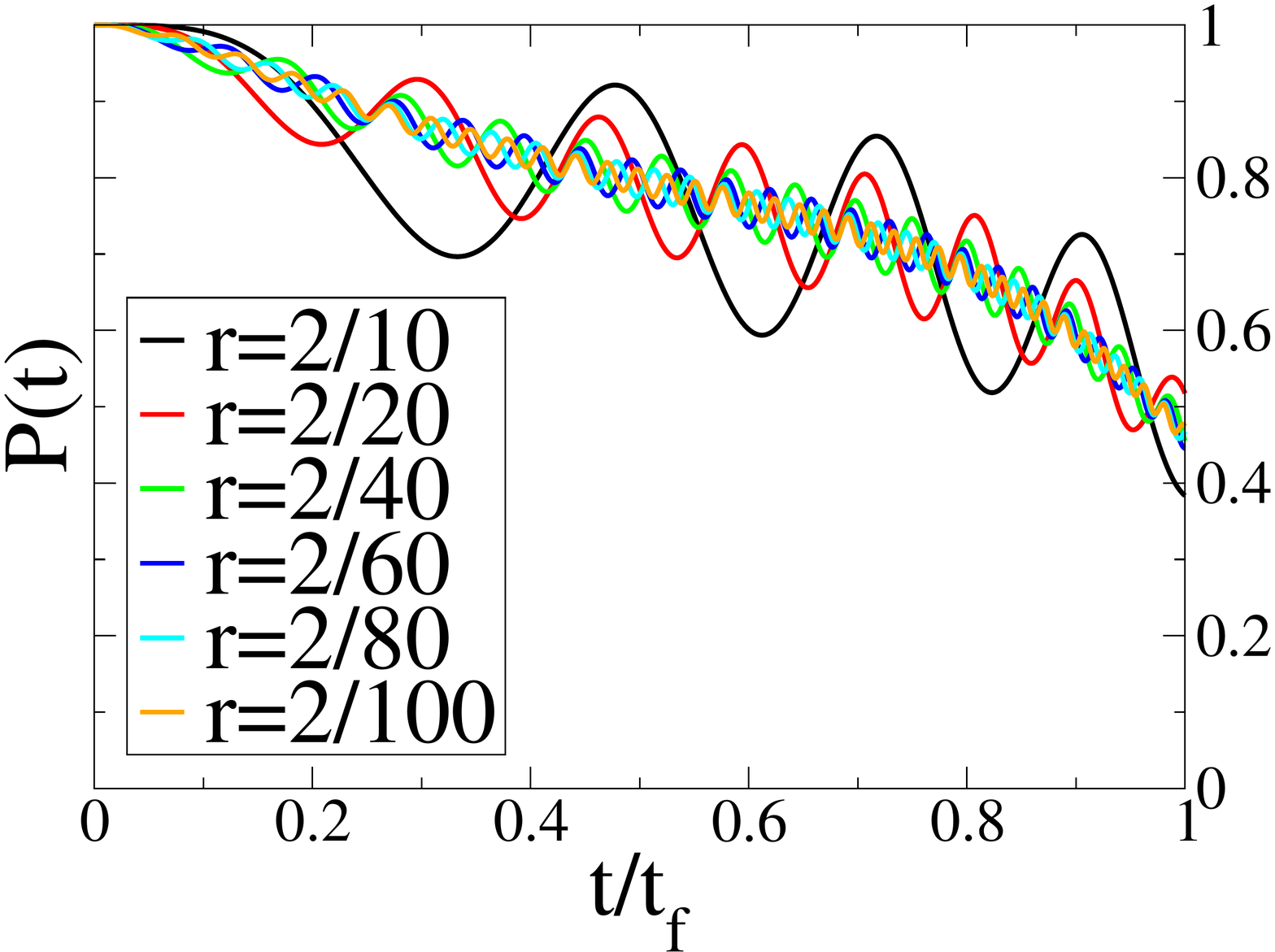}
\includegraphics[width=0.49\columnwidth]{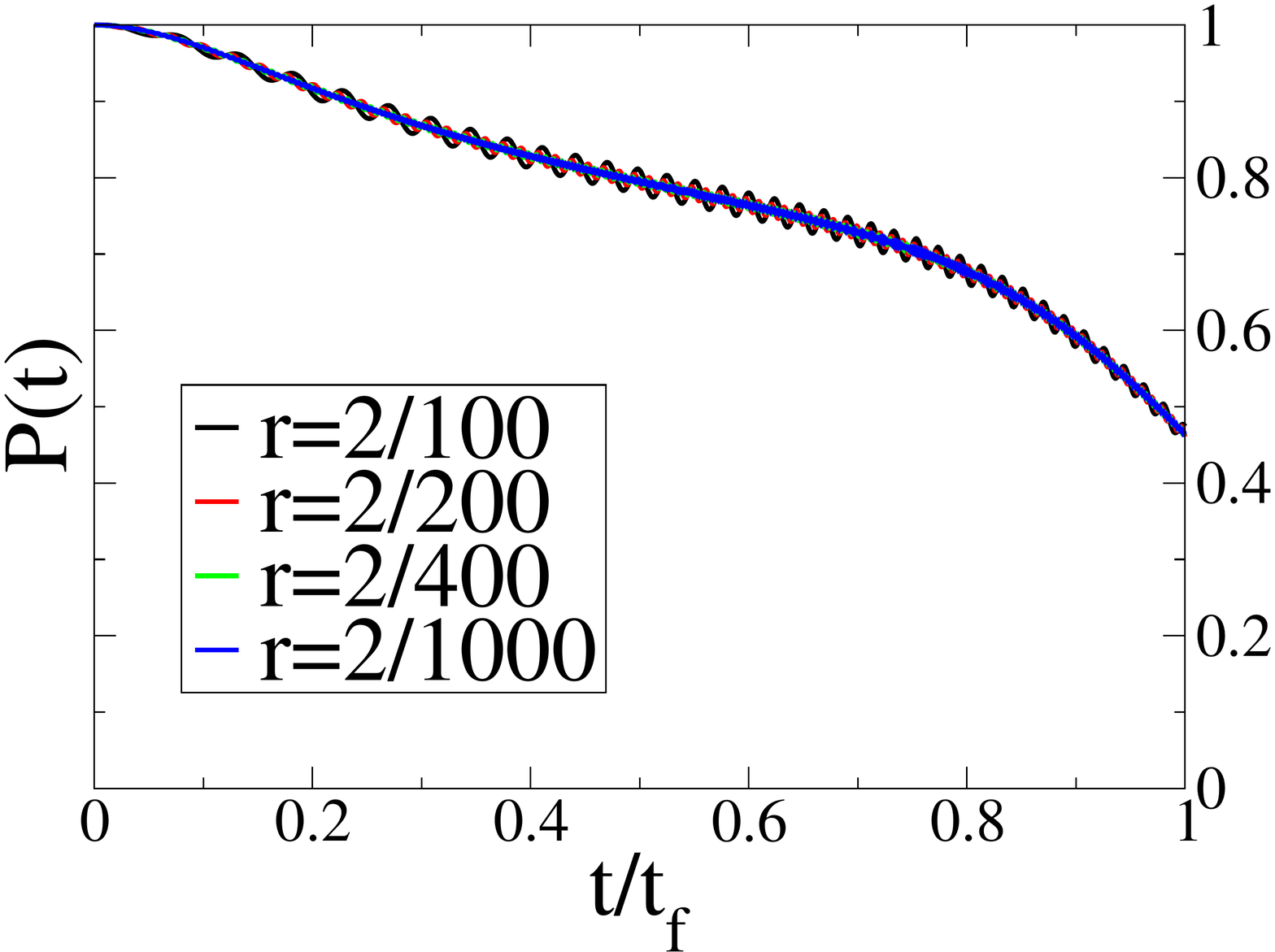}
\caption{\label{fig65}
(Color online)
Time evolution of $P(t)$ for cuts I (upper row) and II (lower row), for different rates with periodic boundary conditions (no edge states).
}
\end{figure}

In Figs. \ref{fig61} we consider the effect of the rate of change on the overlap for cuts $I$ and $II$.
We vary the rates between $r=2/10$ and $r=2/1000$, from relatively fast decays to quite slow parameter changes.
In the case of the first cut, the Majorana decays as expected, but if the decay is relatively fast
($r=2/10$), there is some revival of the overlap after the quantum critical point. These oscillations
decrease as the decay rate becomes smaller. For slow rates these oscillations are vanishingly small.
In the case of the second cut to the $Z_2$ phase, we also see that the fast rates have a behavior that
is consistent with the results obtained for the abrupt quenches. The overlap vanishes and oscillates 
after the quantum critical point. These oscillations decrease in amplitude as the rate becomes smaller,
the mean value becomes finite, particularly for smaller rates, and the overlap is clearly finite. Note that
time is scaled by $t_f$ in order to compare the various decay rates.

In Fig. \ref{fig65} we consider a system with no edge states, by taking periodic boundary conditions
both along the $x$ direction and the $y$ direction. The results presented are the overlap to the
lowest energy state (with finite energy) also at $k_x=0$. As time evolves and for the various
decay rates, the overlap decreases due to the increased distinguishability of the states, but the overlap
remains finite. Note that there are also quantum critical points since the energy gap vanishes and
opens again. However, these are not edge states due to the different boundary conditions, even though
the phases have different Chern numbers. The lowest energy state is therefore more robust and has a
finite overlap with its time evolved single particle state.

\subsection{Defect production}

\begin{figure}[t]
\includegraphics[width=0.95\columnwidth]{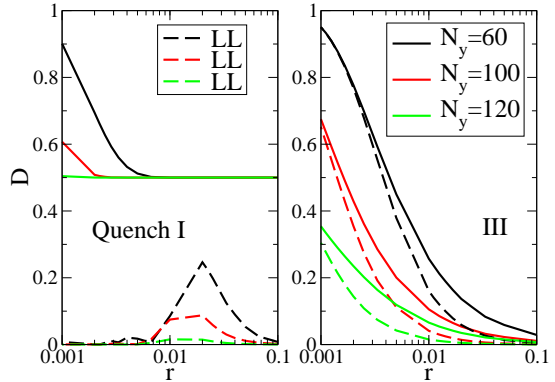}
\caption{\label{fig19}
(Color online)
Defect production after quenches $I$ and $III$ and overlap to the lowest excited state.
}
\end{figure}

When crossing a quantum critical point the Kibble-Zurek mechanism predicts scaling behavior, associated with
the critical slowing down and the appearance of domains of increasing size of the more stable phase,
that scale in a universal way with the rate of parameter change across the phase transition.
In particular, it is expected that the defect production induced by the coupling to excited
states should also scale. In this work we are considering the admixture of excited states
to a Majorana mode as the slow quench is completed.
The defect production is defined as \cite{polkovnikov2,bermudez1}
\be
D=\sum_{\epsilon_n(\xi(t_f))>0} |\langle \psi_n(\xi(t_f)|\psi_0(t_f)\rangle |^2 
\ee
as the sum of the square of the overlaps of the time evolved Majorana state on the
positive energy eigenstates of the Hamiltonian at the final time, $t_f$.
We consider, as examples, a Majorana mode at momentum $k_x=0$ and calculate the defect production for 
the quenches $I,III$, chosen before.

In Fig. (\ref{fig19}) the defect productions are plotted as a function of the rate of
change across the quantum critical points. Also, the contribution from the lowest excited
state is singled out. In the left panel we consider quench $I$ from a topological phase to a 
trivial phase.
For quench I ($C=1 \rightarrow C=0$) we find that the defect production saturates to $1/2$.
This resul is reminiscent of the result
found before for the one-dimensional spinless Kitaev model \cite{bermudez2}.
This result is consistent with the loss of robustness of the Majorana fermion as the transition
occurs. At both low and high rates the overlap to the lowest energy state is small and only
at intermediate rates is significative. Note however, that this lowest energy excited state has
finite energy.
For quench III ($C=-1 \rightarrow C=-2$) the defect production 
has a rather different behavior (similar results are obtained, for instance, for a transition
between $C=-1$ and $C=1$). At small rates the defect production is high and it decreases monotonically
as the rate increases. This result is consistent with the decay of the overlap to the initial Majorana
state as the rate decreases. As discussed above, if the decay rate is small the overlap to the initial
Majorana state is also small and we expect a larger defect production. The results for the contribution
of the lowest excited state show that its weight is quite large. For small rates it basically saturates
the defect production and deceases as the rate increases. Note that in this quench, the lowest energy
excitation is also a Majorana fermion. 
The results for both quenches show the nonuniversal behavior of the Majorana fermion, as obtained for
the Kitaev model.

\section{Conclusions}

In this work the robustness of the Majorana fermions and of the Chern number of
topological phases in a triplet superconductor have been determined.

In general, in the case of a strong spin-orbit coupling system,
a quantum quench leads to a decay of the Majorana modes and to a revival
time that scales with the system size. In some cases these modes are, however, somewhat 
robust. This is particularly observed when a quench connects states in two phases
that share edge states at the same momentum value. 
When the spin orbit-couling is not strong, and the pairing vector is not aligned with
the spin orbit vector, such as in the weak coupling case considered here, the Majorana
modes are more robust, and a finite survival probability is found due to the large
overlaps between the single-particle eigenstates of the two Hamiltonians, the initial
one and the final one. We also found that a signature of the topological phase,
the Chern number, remains unchanged after the quench until the propagating time-evolved
Majorana state reaches a peak at the center of the system, beyond which the Chern number
fluctuates increasingly. It was also found that, in some cases, the time averaged Chern number
seems to converge to the value expected of the final state, but in most cases this convergence
is very slow, if it converges at all.

The results for slow quenches lead to similar conclusions and show that the Kibble-Zurek scaling
does not hold for the decay of the Majorana modes, as found in other topological edge states.

\section*{Acknowledgements}

The author thanks discussions with Pedro Ribeiro and Antonio Garcia-Garcia and partial
support by the Portuguese FCT under grant PEST-OE/FIS/UI0091/2011 and 
grant PTDC/FIS/111348/2009.

\end{document}